\documentclass[12pt]{iopart}

\usepackage{graphicx}% Include figure files
\usepackage{citesort}
\usepackage{amssymb}
\usepackage{color}
\usepackage{multirow}

\begin{document}

\title[Electron heating via the plasma series resonance in capacitive plasmas]{Electron heating via the self-excited plasma series resonance in geometrically symmetric multi-frequency capacitive plasmas}

\author{E. Sch\"ungel$^1$, S. Brandt$^1$, Z. Donk\'o$^2$, I. Korolov$^2$, A. Derzsi$^2$, J. Schulze$^1$}
\address{$^1$Department of Physics, West Virginia University, Morgantown, West Virginia 26506-6315, USA \\
$^2$Institute for Solid State Physics and Optics, Wigner Research Centre for Physics, Hungarian Academy of Sciences, 1121 Budapest, Konkoly-Thege Mikl\'os str. 29-33, Hungary}

\begin{abstract}

The self-excitation of Plasma Series Resonance (PSR) oscillations plays an important role in the electron heating dynamics in Capacitively Coupled Radio Frequency (CCRF) plasmas. In a combined approach of PIC/MCC simulations and a theoretical model based on an equivalent circuit, we investigate the self-excitation of PSR oscillations and their effect on the electron heating in geometrically symmetric CCRF plasmas driven by multiple consecutive harmonics. The discharge symmetry is controlled via the Electrical Asymmetry Effect, i.e. by varying the total number of harmonics and tuning the phase shifts between them. It is demonstrated that PSR oscillations will be self-excited under both symmetric and asymmetric conditions, if (i) the charge-voltage relation of the plasma sheaths deviates from a simple quadratic behavior and if (ii) the inductance of the plasma bulk exhibits a temporal modulation. These two effects have been neglected up to now, but we show that they must be included in the model in order to properly describe the nonlinear series resonance circuit and reproduce the self-excitation of PSR oscillations, which are observed in the electron current density resulting from simulations of geometrically symmetric CCRF plasmas. Furthermore, the effect of the PSR self-excitation on the discharge current and the plasma properties, such as the potential profile, is illustrated by applying Fourier analysis. High frequency oscillations in the entire spectrum between the applied frequencies and the local electron plasma frequency are observed.
As a consequence, the electron heating is strongly enhanced by the presence of PSR oscillations. A complex electron heating dynamics is found during the expansion phase of the  sheath, which is fully collapsed, when the PSR is initially self-excited. The Nonlinear Electron Resonance Heating associated with the PSR oscillations causes a spatial asymmetry in the electron heating. By discussing the resulting ionization profile in the non-local regime of low-pressure CCRF plasmas, we examine why the ion flux at both electrodes remains approximately constant, independently of the phase shifts.

\end{abstract}

\maketitle 

\section{Introduction}

The question how electrons are heated in Capacitively Coupled Radio Frequency (CCRF) plasmas is of fundamental importance, yet in most cases the answer is  complicated and diverse. Typically, the electron heating dynamics reveals a complex behavior in both space and time. The most prominent mechanisms are the interactions of electrons with the high electric field in the plasma sheath regions, which oscillates periodically within the RF period. 
At low pressures electrons primarily gain energy near the edge of the expanding sheath ($\alpha$-mode heating \cite{Belenguer,Fermi,Surendra,Gozadinos,Salabas,Vender,Schulze_dual,Tochikubo,Schulze_eheat,Dittmann,Kuellig,Killer,Gans,Mahony,Schulze_field_reversal,UCZ_field_reversal,ELIAS1,Perrin,Boeuf,BulkMode,BulkMode3,Goedheer,PSR_beams_Schulze,Schulze_eheat_IEEE,ambipolar}). Secondary electrons released from the electrode surface gain high energies around the time of maximum sheath extension ($\gamma$-mode heating \cite{Belenguer,Schulze_dual,Tochikubo,Schulze_eheat,Dittmann,Kuellig,Gans,Mahony,BulkMode,BulkMode3}). Under certain conditions, an additional source of heating is provided by a local reversal of the electric field at times when the sheaths collapse \cite{Gans,Mahony,Schulze_field_reversal,UCZ_field_reversal}. In dusty and electronegative plasmas, as well as in CCRF plasmas operated at high pressures the electron heating by drift and/or ambipolar fields within the plasma bulk contributes dominantly to the overall energy gain of electrons \cite{Belenguer,Killer,ELIAS1,Perrin,Boeuf,BulkMode,BulkMode3,Goedheer,Kersten}. 
The electron heating in the $\alpha$-mode consists of a collisionless (stochastic) and a collisional (Ohmic) component. Stochastic heating refers to the interaction of electrons with the electric field at the plasma sheath edge \cite{Tatanova}, whereas Ohmic heating refers to the change of electron momentum in collisions. A detailed discussion on these terms can be found in a recent review by Lafleur et al. \cite{Lafleur_heating}. 

Moreover, resonance phenomena can significantly affect the total electron heating. Oscillations in the discharge current at resonance frequencies above the applied frequency can be self-excited in CCRF plasmas due to their nonlinear electrical characteristics. In an equivalent circuit, the electron inertia corresponds to an inductance of the plasma bulk, which forms a parallel circuit with the bulk capacitance and a series circuit with the nonlinear sheath capacitances. Generally, both parallel and series resonances may occur \cite{PSR_Allen1,PSR_Allen2,PSR_Allen3}. In low pressure CCRF plasmas the plasma series resonance (PSR) is, however, typically the more pronounced resonance feature. Many studies have shown that the electron heating associated with the self-excitation of high frequency PSR oscillations strongly enhances the total electron heating \cite{PSR_Allen1,PSR_Allen2,PSR_Allen3,PSR_Klick,PSR_Birdsall1,PSR_Mussenbrock,PSR_nonlinear_Mussenbrock,PSR_Semmler,PSR_diagnostics_Schulze,PSR_beams_Schulze,PSR_Ziegler1,PSR_stoch_Schulze,PSR_Lieberman1,PSR_Lieberman2,PSR_EAE_Zoltan,PSR_Yamazawa,EAE3,PSR_Ziegler2,PSR_Wang,PSR_Bora3,PSR_Bora4,PSR_Bora1,PSR_Bora2,PSR_Birdsall2,PSR_model_UCZ}. Under the conditions considered by \textit{Ziegler et al.} \cite{PSR_Ziegler1}, for instance, the resonance heating accounted for about one half of the total electron energy gain. In a kinetic description of heating mechanisms in CCRF plasmas, the phase resolved optical emission spectroscopy (PROES) measurements of \textit{Schulze et al.} \cite{PSR_diagnostics_Schulze,PSR_beams_Schulze,PSR_stoch_Schulze} verified the generation of multiple beams of highly energetic electrons during a single phase of sheath expansion by the stepwise sheath expansion in the presence of the PSR. \textit{Klick et al.} demonstrated that the detection of the PSR oscillations by a current sensor implemented into the grounded chamber wall allows determining the electron density and temperature via Self-Excited Electron Resonance Spectroscopy (SEERS) \cite{PSR_Klick,PSR_SEERS_Klick}. Self-excited oscillations are not limited to CCRF plasmas, but can be found in various other types of bounded plasmas, such as inductively coupled plasmas operated in the E-mode \cite{PSR_ICP_Kempkes,PSR_ICP_Boffard} or hybrid combinations of inductively/capacitively coupled plasmas \cite{PSR_ICPCCP_Lee,PSR_ICP_coupling}, too.

A detailed understanding of each of the individual mechanisms contributing to the electron heating dynamics is the basis for controlling and improving these plasmas for the broad variety of their applications, such as the etching of nanometer size structures in the manufacturing of integrated circuits \cite{Makabe,Donnelly}, the deposition of nano structures \cite{nano} as well as thin films or functional coatings \cite{Rath,Hrunski,Micromorph}, and the surface modification of polymers for biomedical applications \cite{biomed1,biomed2}. For all these applications, a precise control of the plasma surface interaction is required. The process rate is governed by the flux-energy distributions of reactive species onto the surface. These distributions, in turn, are the result of various energy dependent processes involving electron collisions within the plasma volume. Hence, they critically depend on the spatio-temporal changes of the electron energy distribution function and, thereby, on the gain and loss of energy of the electrons.

Large efforts have been invested in the last decades to diagnose, control, and optimize CCRF plasmas. A major step has been the application of two and more sources with different frequencies to the powered electrode. In the "classical" approach, the applied voltage has two components: a low-frequency component, which is used to control the mean sheath voltage and, thereby, the ion energy at the electrode, and a high-frequency component, which is used to control the electron heating due to the higher efficiency at higher frequencies \cite{dual1,dual2,dual3,dual4,PSR_Mussenbrock,fundamental}. Although this approach has led to strong improvements, the control is not ideal due to the effects of frequency coupling and secondary electrons \cite{fcoupling1,fcoupling2,fcoupling3,secondaries,fundamental}. Since 2008, an alternative approach, called Electrical Asymmetry Effect (EAE) \cite{PSR_EAE_Zoltan,PSR_EAE_geomasymm,EAE2,EAE3,EAE5,EAE7,EAE8,EAE12,EAE13,EAE15,EAE16,EAEmultif1,EAEmultif2,EAEjohnson,EAEjohnson2,EAEjohnson3,VWT_Lafleur,EAEbienholz,ELIAS2,ELIAS3,EAE_IRLAS_DEPO,fundamental,Diomede1}, has been investigated. In contrast to the superposition of two frequencies separated by about one order of magnitude in the "classical" dual-frequency approach, here the discharge is operated at a combination of a fundamental frequency with one or more of its subsequent harmonics. By tuning the phase shifts between these harmonics, the symmetry of the applied voltage waveform can be varied. This, in turn, allows for a control of the DC self-bias, the time averaged voltage drops across the powered and grounded electrode sheaths, and, thereby, of the mean ion energy at both electrodes. (Note that a controllable DC self-bias also appears in this case in geometrically symmetrical reactors.) At the same time, the ion flux remains approximately constant, because the applied voltage amplitude is kept constant and the dependence of the overall electron heating on the phase shift is weak under most conditions \cite{EAE12}. Thus, the EAE provides a convenient way of controlling the ion energy independently of the ion flux. This concept has been tested in dual-frequency experiments \cite{EAE5,EAE12,PSR_EAE_geomasymm,ELIAS1,EAEbienholz,EAE_IRLAS_DEPO,ELIAS3,EAE13,EAE16} and simulations \cite{PSR_EAE_Zoltan,PSR_EAE_geomasymm,EAE2,EAE3,EAE8,EAE12,EAE15,ELIAS2,ELIAS3,EAE16} and has proven to be beneficial for surface processing applications of CCRF plasmas \cite{Hrunski,EAEbienholz,EAE_IRLAS_DEPO}. The control range of the EAE can be extended by using more than two frequencies \cite{EAEmultif1,EAEmultif2}, also called voltage waveform tailoring (VWT) \cite{EAEjohnson,EAEjohnson2,EAEjohnson3,VWT_Lafleur,Diomede1}. Different from the approach of arbitrary substrate biasing in remote plasma sources \cite{Wendt1,Wendt2,Baloniak,Diomede2}, the discharge is sustained by the customized voltage waveform.

The shape of the voltage waveform affects the electron heating in multiple ways. Asymmetries in the electron heating and subsequent excitation dynamics have been observed in geometrically symmetric, electrically asymmetric dual-frequency CCRF discharges due to the asymmetry of the applied voltage waveform at low pressures and the discharge current at high pressures, respectively \cite{EAE7}. \textit{Derzsi et al.} have shown that the electron heating is strongly enhanced by adding more consecutive harmonics to the applied voltage waveform and that separate control of ion energy and ion flux is hardly possible in the $\gamma$-mode \cite{EAEmultif2}. Under conditions explored by \textit{Lafleur et al.}, it is even possible to customize the electron heating at each of the two sheaths, leading to a control of the ion flux at either side of the discharge, without a strong variation of the ion energy \cite{Lafleur_ionflux1,VWT_Lafleur}. Furthermore, the application of sawtooth-shaped voltage waveforms causes different expansion velocities of the powered and grounded electrode sheaths and the temporal asymmetry results in an asymmetry of the discharge via the asymmetric heating profile of electrons in the $\alpha$-mode \cite{Lafleur_asymm1,Lafleur_asymm2}. 

A paper by \textit{Donk\'o et al.} has, however, demonstrated that self-excited PSR oscillations may occur in geometrically symmetric discharges when operated using asymmetric dual-frequency voltage waveforms \cite{PSR_EAE_Zoltan}. 
%In agreement with the studies on the PSR in asymmetric CCRF plasmas mentioned above, the electron heating is found to be enforced by the presence of the PSR. However, claiming the discharge asymmetry to be a necessary condition for the self-excitation of PSR oscillations would be incorrect:
As a further step, very recent investigations by \textit{Sch\"ungel et al.} revealed that the self-excitation of the PSR is possible even in symmetric single-frequency capacitive plasmas. The requirement for this is that the resonance circuit is nonlinear; this is well possible in asymmetric as well as symmetric multi-frequency discharges \cite{PSR_EAE_Letter}. This finding is important, since the self-excitation of the PSR was purely attributed to the presence of a discharge asymmetry in previous works.

In this paper we study the electron heating associated with the self-excitation of plasma series resonance oscillations in geometrically symmetric, electrically asymmetric CCRF plasmas. We demonstrate that a discharge asymmetry is not the only reason for the self-excitation of the PSR. In particular, this work aims at answering the following questions: What causes the self-excitation in geometrically symmetric configurations and how can these mechanisms be described theoretically? What is the effect of the PSR on the electron heating for different numbers of applied voltage harmonics? How does it affect the symmetry of the heating and, eventually, the symmetry of the particle densities and fluxes?
We investigate these issues using both self-consistent, kinetic simulations and an equivalent circuit model. 
The structure of this paper is as follows: The basics of the simulations and the model are described in the next chapter. After that, the results are presented in chapter 3 in two parts, as we first discuss the self-excitation of the PSR and then examine the role of the PSR for the electron heating and subsequent ionization dynamics. Finally, conclusions are drawn in chapter 4.

\section{Simulation and model}
\subsection{PIC/MCC Simulation}

\begin{figure}[tbp]
\centering
\includegraphics[width=0.65\textwidth]{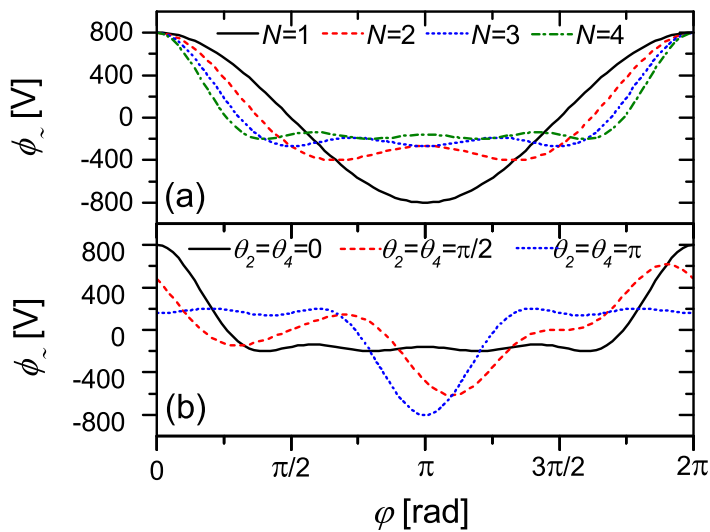}
\caption{Applied voltage waveform as a function of the RF phase. (a) Different total number of harmonics: $N=$ 1 (black solid line), $N=$ 2 (red dashed line), $N=$ 3 (blue dotted line), $N=$ 4 (green dash-dotted line). All phases are set to zero. (b) Different phase shifts for $N=$ 4: $\theta_2=0$ and $\theta_4=0$ (black solid line), $\theta_2=\pi/2$ and $\theta_4=\pi/2$ (red dashed line), $\theta_2=\pi$ and $\theta_4=\pi$ (blue dotted line). $\theta_1$ and $\theta_3$ are kept zero.} 
\label{fig1}
\end{figure}

Our plasma simulations are based on the Particle In Cell technique complemented with a Monte Carlo treatment of collisions (PIC/MCC). The code is one-dimensional in space and three-dimensional in velocity space. About $10^5$ superparticles represent each type of charged species (electrons and ions). We consider a symmetrical geometry: two planar, parallel, and infinite electrodes separated by a distance of $d=$ 30 mm. The powered electrode is located at $z=$ 0 and the grounded electrode is located at $z=d$.
% is advantageous for a fundamental investigation of the PSR self-excitation, because in more complicated two- or three-dimensional configurations a complicated multi-mode ansatz has to be chosen \cite{PSR_nonlinear_Mussenbrock}. 
We simulate argon plasmas using the cross section data provided by \textit{Phelps} \cite{PhelpsOnline,Phelps1,Phelps2}. The neutral background gas is at a pressure of 3 Pa and a temperature of 400 K. The applied voltage waveform, $\phi_{\sim}(\varphi)$, is defined by
\begin{eqnarray}
\phi_{\sim}(\varphi) &= & \phi_{tot}  \sum_{k=1}^N \frac{2(N-k+1)}{N(N+1)} \textnormal{cos}(k \varphi + \theta_k) \\
&=& \sum_{k=1}^N \phi_k \textnormal{cos}(k \varphi + \theta_k),
\label{AppVol}
\end{eqnarray}
where $\phi_{tot}$ is the total voltage amplitude and $N$ is the number of applied harmonics (see figure \ref{fig1}). Here, $\phi_{tot}=800$ V. Each frequency has an individual phase, $\theta_k$, and amplitude, $\phi_k$, which is chosen based on the criterion introduced in Ref. \cite{EAEmultif1}. The criterion ensures the strongest asymmetry of the applied voltage waveform. 
There are two types of waveform asymmetries: a temporal asymmetry, i.e. $\phi_{\sim}(\varphi) \neq \phi_{\sim}(-\varphi)$, and an electrical (amplitude) asymmetry, i.e. $\phi_{\sim,max} \neq -\phi_{\sim,min}$. The voltage waveforms with all phases set to zero, which are in the focus of this paper, are temporally symmetric but electrically asymmetric due to the difference between the maximum and the absolute value of the minimum. Generally, when $\theta_k$ is varied, there can be a temporal asymmetry, too.
Tuning the phase shifts of the even harmonics allows for decreasing and reversing the asymmetry. $\varphi=2 \pi f t$ is the RF phase with  $f=13.56$ MHz being the fundamental frequency.  Both the electron reflection coefficient and the $\gamma$-coefficient for ion induced secondary electron emission are taken to be 0.2 at both electrodes.

In the analysis of the spatially and temporally resolved simulation data, the momentary plasma sheath edge is calculated based on the integral criterion introduced by Brinkmann \cite{Brinkmann_criterion}, where the electron density inside the sheath is balanced by the net charge on the bulk side
\begin{equation}
\int_0^{s(\varphi)} n_e(z) dz = \int_{s(\varphi)}^{d/2} \left[ n_i(z) - n_e(z) \right] dz .
\end{equation}
The electron current density is determined from the spatially and temporally resolved electron current by averaging over a region within the plasma bulk, which remains quasineutral at all times within the RF period. The electron heating rate, $P_e(z,\varphi)$, is obtained in the PIC/MCC simulations by adding up the local gain and loss of energy of electrons at all positions in space and time. The heating will be compared to the results of a model, which does not include any cooling of electrons. Therefore, we consider only positive values and define
\begin{equation}
P_{e}^{+}(z,\varphi)=P_e(z,\varphi) \, \Theta\left[ P_e(z,\varphi) \right],
\end{equation}
where $\Theta$ is the Heaviside step function. Then, the electron heating accumulated within the RF period is determined via integration over the RF phase. Moreover, we split the discharge in space into the half adjacent to the powered electrode and the half adjacent to the grounded electrode. Spatial integration over the respective discharge half yields an accumulated electron heating in either half:
\begin{eqnarray}
\label{Pep}
\int_0^\varphi P_{e,p}^{+} (\varphi') d\varphi' &=&\int_0^{d/2} \int_0^\varphi P_{e}^{+}(z,\varphi') d\varphi' dz, \\
\int_0^\varphi P_{e,g}^{+} (\varphi') d\varphi' &=&\int_{d/2}^{d} \int_0^\varphi P_{e}^{+}(z,\varphi') d\varphi' dz.
\label{Peg}
\end{eqnarray}

For the spectral analysis we record at points ($z_u$, $u=1,2,...,1024$) of the spatial grid used in the simulation and at discrete times $t_s = s \Delta t$ ($s = 1,...,M$) the values of the electron density $n_e(z_u,t_s)$, the electric field $E(z_u,t_s)$, and the potential $\phi(z_u,t_s)$. The length of the time series, $M$, spans 8 RF cycles with 32,768 time steps within each RF cycle, resulting in $M = 262,144$. Taking the potential as an example, the power density spectrum is calculated via discrete Fourier transform as: 
\begin{equation}
\label{EqPSD}
\hat{\phi}(z,\omega) = | \mathcal{F} \{ \phi(z,t) \} |^2   % may need different notations !!
\end{equation} 
Note that the notation for the discrete character of $z$, $t$, and $\omega$ was omitted here. Prior to the Fourier transform we apply a Hann window function to minimize effects originating from the finite length of the time series. The spectra of the different quantities mentioned above are calculated after each 8 RF cycles and the results are averaged within one simulation, and finally for independent runs, to improve the signal to noise ratio. The resolution of the spectra with the parameter values specified above is, e.g. $\omega/\omega_{\rm p} \approx 2 \times 10^{-3}$ for the single harmonic, 13.56 MHz excitation.

%Finally, we visualize the effect of the PSR on the plasma dynamics by calculating the power spectral density, 
%\begin{equation}
%S_E(z,\omega)= \left| \frac{1}{T} \int_0^T e^{-i \omega t}  \phi(z,t) dt  \right|^2, \, Zoli?
%S_E(z,\omega)= \frac{1}{T} \int_0^T e^{-i \omega t}\phi(z,t) dt   \int_0^T e^{i \omega t}\vec{E}^{\ast}(z,t) dt
%\end{equation}
%of the electric potential. For this purpose a Fourier transformation of the absolute square of the spatio-temporal potential distribution, $\phi(z,t)$, is performed with a high sampling rate of $2^{14}$ points per RF period (corresponding to a temporal resolution of about 4.5 ps). The noise is reduced by averaging over about $10^4$ RF cycles.

\subsection{Model}

\begin{figure}[tbp]
\centering
\includegraphics[width=0.65\textwidth]{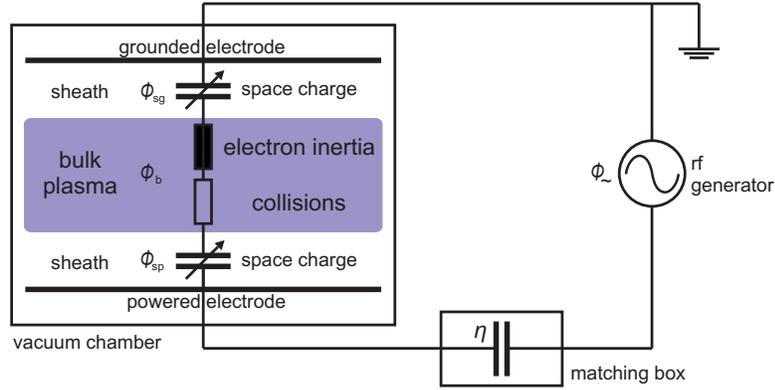}
\caption{Schematical view of the different components of an electrical circuit of a CCRF discharge including the equivalent circuit components of the plasma bulk and sheaths.} 
\label{figCCRFcircuit}
\end{figure}

In this section, a differential equation for the normalized charge in the powered electrode sheath, $q$, describing PSR oscillations is derived from an equivalent circuit model by formulating all internal voltages as functions of $q$. The series resonance circuit is formed by the sheath capacitances, the resistance of the bulk due to collisions, and its inductance due to electron inertia, respectively (see figure \ref{figCCRFcircuit}). Then, it will be explained how the electron current and the electron heating are determined.

Following Kirchhoff's rule for closed loops, the voltage balance of a CCRF discharge is obtained as \cite{PSR_model_UCZ,EAE2,EAE13,ELIAS1,PSR_EAE_Letter}
\begin{equation}
\label{VolBal}
\eta+\phi_{\sim}(\varphi)=\phi_{sp}+\phi_b+\phi_{sg},
\end{equation}
i.e. the sum of the applied voltage waveform, $\phi_{\sim}$, and the DC self-bias, $\eta$, must equal the sum of the individual voltage drops across the powered electrode sheath, $\phi_{sp}$, the plasma bulk region, $\phi_b$, and the grounded electrode sheath, $\phi_{sg}$. It can be assumed that the sheath regions are free of electrons, as the mean electron energy is much smaller than the voltage drop across the sheaths. Hence, integrating Poisson's equation between the electrode (at $z=$ 0) and the momentary plasma sheath edge (at $z=s(\varphi)$) yields
\begin{eqnarray}
Q_{sp}(\varphi)&=&e \int_0^{s(\varphi)} n_i(z) dz, \\
\phi_{sp}(\varphi)&=&-\frac{e}{\epsilon_0} \int_0^s \int_z^{s(\varphi)} n_i(z') dz' dz
\end{eqnarray}
for the charge (over a unit surface area), $Q_{sp}$, within the powered electrode sheath and the voltage drop across this sheath. Here, $e$ and $\epsilon_0$ are the elementary charge and the vacuum permittivity, respectively. Any temporal modulation of the ion density profile, $n_i(z)$, is neglected. For a matrix sheath, the ion density is homogeneous, $n_i(z)=n_{i0}$, and the charge-voltage relation is simply found as $\phi_{sp}=-Q^2/(2e \epsilon_0 n_{i0})$. However, this quadratic approximation corresponds to a strong simplification; in fact, we will see that the matrix sheath is an over-simplified picture, which is not capable of reproducing the real physical sheath behavior. 

As a more realistic charge-voltage relation, we adopt the functional relation proposed in reference \cite{UCZ_sheath_model}:
\begin{equation}
\label{Usp}
\bar{\phi}_{sp}(\varphi) = -q_{tot}^2 q^2(\varphi) \left[ q(\varphi) (1-a) +a \right].
\end{equation}
Here, $\bar{\phi}_{sp}(\varphi)= \phi_{sp}(\varphi)/\phi_{tot}$ is the normalized powered sheath voltage and $q_{tot}$ is the normalized total net charge within the discharge volume, which is assumed to be constant \cite{EAE2,EAE8,EAE13}. Thus, $q$ varies between $q=0$, i.e. any residual voltage or charge of the sheath at the time of collapse is neglected, and $q=1$ corresponding to a fully expanded sheath. This cubic ansatz (equation (\ref{Usp})) is motivated by a power series \cite{UCZ_sheath_model}. It requires another parameter, $a$, which depends on the ion density profile in the sheath region: $a=1$ corresponds to the quadratic behavior of a matrix sheath, whereas $a>1$ allows for a cubic contribution. Similarly, one can find a cubic relation for the charge-voltage relation of the grounded electrode sheath. Here, we make use of the assumption of a constant total charge, so that the normalized charge in the grounded sheath is $1-q$. Hence \cite{UCZ_sheath_model},
\begin{equation}
\label{Usg}
\bar{\phi}_{sg}(\varphi) = \varepsilon q_{tot}^2 \left[ 1- q(\varphi) \right]^2 \left[ [1-q(\varphi)] (1-b) +b \right].
\end{equation}
The parameter $b$ in equation (\ref{Usg}) has the same meaning as the parameter $a$ in equation (\ref{Usp}). The powered and grounded sheath voltages are linked via the so-called symmetry parameter, \cite{EAE2,EAE8,EAE13}
\begin{equation}
\label{SymPar}
\varepsilon = \left| \frac{\phi^{max}_{sg}}{\phi^{max}_{sp}} \right| \approx \left( \frac{A_p}{A_g} \right)^2 \frac{\bar{n}_{sp}}{\bar{n}_{sg}},
\end{equation}
which is very important for the model description of PSR oscillations. $A_p$ and $A_g$ are the surface areas of the powered and grounded surface areas and $\bar{n}_{sp}$ and $\bar{n}_{sg}$ are the mean ion densities in the respective sheaths.

$\varepsilon=1$ is the criterion for a symmetric discharge, whereas $\varepsilon \neq1$ is the criterion for an asymmetric one. In the latter case, this means that the maximum sheath voltages are different, which is caused by a difference in the ion density profiles adjacent to both electrodes under most conditions. Note that the presence of a large DC self-bias amplitude, which occurs in geometrically asymmetric discharges, also corresponds to an asymmetry (of the potential profile). However, it does not necessarily correspond to an asymmetry of the discharge itself, i.e. an asymmetry of the density profiles. On the other hand, a discharge can be asymmetric ($\varepsilon \neq 1$), although the DC self-bias vanishes \cite{PSR_EAE_geomasymm}. Thus, $\varepsilon$ is regarded to be the quantity defining the ``discharge asymmetry''. 

Almost all studies on the self-excitation of PSR oscillations investigated discharges, which were inherently strongly asymmetric due to the predefined chamber configuration: typically, a small powered electrode in a vacuum vessel exhibiting a relatively large grounded surface area was considered, i.e. $A_p \ll A_g$. Moreover, usually a quadratic charge-voltage relation for the RF sheaths was assumed and PSR oscillations were found to be self-excited only in asymmetric discharges ($\varepsilon \neq 1$), since otherwise ($\varepsilon =1$) the quadratic nonlinearity in the voltage balance cancelled. The multi-frequency approach provides a more versatile scenario: for instance, a geometrically symmetric parallel plate discharge can be operated by an asymmetric voltage waveform. This causes a deviation of the symmetry parameter from unity, e.g. $\varepsilon \approx 0.5$ in Ref. \cite{PSR_EAE_Zoltan}. Moreover, recent investigations have indicated that the PSR oscillations can occur in symmetric ($\varepsilon = 1$) CCRF plasmas, as well \cite{PSR_EAE_Letter}.

\begin{figure}[tbp]
\centering
\includegraphics[width=0.65\textwidth]{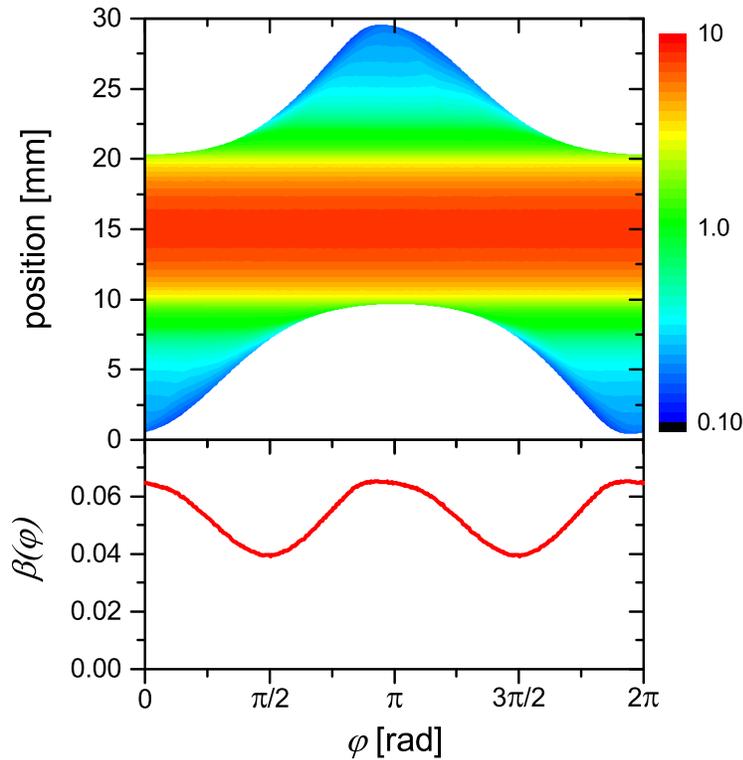}
\caption{Top: spatio-temporal distribution of the electron density within the plasma bulk (between the plasma sheath edges) obtained from a single-frequency PIC/MCC simulation. The logarithmic color scale gives density values in $10^{15}$ m$^{-3}$. Bottom: bulk parameter $\beta (\varphi)$ within one RF period calculated from the simulation data. Discharge conditions: Ar, 3 Pa, $d=$ 30 mm, $\phi_{tot}=800$ V.}
\label{fig22}
\end{figure}

The plasma bulk is assumed to be a quasineutral region in which the current is purely conductive, i.e. the RF current is carried by the ensemble of bulk electrons. This conduction current can be described based on the momentum balance equation,
\begin{equation}
\label{MomBal}
m n \frac{\partial u}{\partial t}=-e n E - m n \nu_m u,
\end{equation}
 where $m$, $n$, $u$ are the electron mass, density, and drift velocity in $z$-direction, i.e. perpendicular to the electrodes. $E$ and $\nu_m$ are the electric field and the electron collision frequency for momentum transfer, respectively. Note that this is a simplified version of the full momentum balance equation, as the convective term on the left hand side and other force densities on the right hand side, such as a pressure gradient, are neglected. We define the electron conduction current density as $j=-e n u$ and neglect any temporal modulations of the electron density when taking the derivative $\partial j/ \partial \varphi$.
Then, solving equation (\ref{MomBal}) yields the electric field in the configuration considered here:
\begin{equation}
\label{BulkField}
E(\varphi)=\frac{m}{e^2 n} \left[ \omega \frac{\partial j (\varphi)}{\partial \varphi} + \nu_m j(\varphi) \right].
\end{equation}
Note that the electric field distribution would be much more complicated in most asymmetric chamber configurations, eventually leading to multiple different current paths and the self-excitation of multiple PSR modes \cite{PSR_nonlinear_Mussenbrock}. However, in the one-dimensional scenario considered here the bulk voltage is simply found by integrating the electric field along the unique current path in the $z$-direction,
\begin{eqnarray}
\label{BulkVol1}
\phi_{b}(\varphi) &=& -\int_{s_p(\varphi)}^{d-s_g(\varphi)} E(\varphi) dz, \\
\bar{\phi}_b (\varphi) &=& - 2 \beta^2(\varphi) \left[ \ddot{q} (\varphi) + \kappa \dot{q} (\varphi)  \right] .
\label{BulkVol2}
\end{eqnarray}
Here, a dot denotes differentiation with respect to the RF phase $\varphi$. Equation (\ref{BulkVol2}) has been normalized by $\phi^{max}_{sp}$. The parameter $\kappa=\nu_m / \omega$ characterizes the collisionality and, hence, corresponds to a damping factor for the PSR. The "bulk parameter"
\begin{equation}
\label{EqBeta}
\beta (\varphi) = \frac{\omega}{\bar{\omega}_{pe} (\varphi)} \sqrt{ \frac{L_b(\varphi)}{s_{max}} }
\end{equation}
is typically small in low pressure electropositive plasmas ($\beta \ll 1$) \cite{PSR_model_UCZ}. It depends on the ratio of the bulk length, $L_b(\varphi)$, to the maximum powered electrode sheath extension, $s_{max}$, and on the spatially averaged inverse electron plasma frequency
\begin{equation}
\bar{\omega}_{pe}^{-1}(\varphi)=\sqrt{ \frac{\epsilon_0 m_e}{e^2 L_{b}(\varphi)} \int_{s_p(\varphi)}^{d-s_g(\varphi)} n_e^{-1}(\varphi,z) dz}.
\end{equation}
In contrast to previous works, we allow for a temporal modulation and explicitly determine $\beta$ as a function of $\varphi$ from the PIC/MCC simulation data. The typical behavior of the electron density profile between the modulated sheaths and the resulting bulk parameter $\beta (\varphi)$ obtained from a single-frequency PIC/MCC simulation is illustrated in figure \ref{fig22}. $\beta$ is maximum at times of sheath collapse at either the powered or the grounded electrode. This is due to the fact that the electron density decreases towards the electrodes, because it follows the ion density that decreases due to the acceleration of ions in the sheath electric field at approximately constant ion flux. Therefore, the local electron plasma frequency decreases accordingly and the inverse effective electron plasma frequency increases at times of sheath collapses, leading to an increase in $\beta(\varphi)$.

The model equation describing the self-excitation of PSR oscillations can now be constructed by inserting equations (\ref{Usp}), (\ref{Usg}), and (\ref{BulkVol2}) into equation (\ref{VolBal}):
\begin{eqnarray}
\label{PSRequation} 
\bar{\eta}+\bar{\phi}_{\sim}(\varphi) = && -q_{tot}^2 q^2(\varphi) \left[ q(\varphi) (1-a) +a \right]  \nonumber \\ 
&& +  \varepsilon q_{tot}^2 \left[ 1- q(\varphi) \right]^2 \left[ [1-q(\varphi)] (1-b) +b \right] \nonumber \\
&& - 2 \beta^2(\varphi) \left[ \ddot{q} (\varphi) + \kappa \dot{q} (\varphi)  \right] .
\end{eqnarray}

\begin{figure}
\centering
\includegraphics[width=0.63\textwidth]{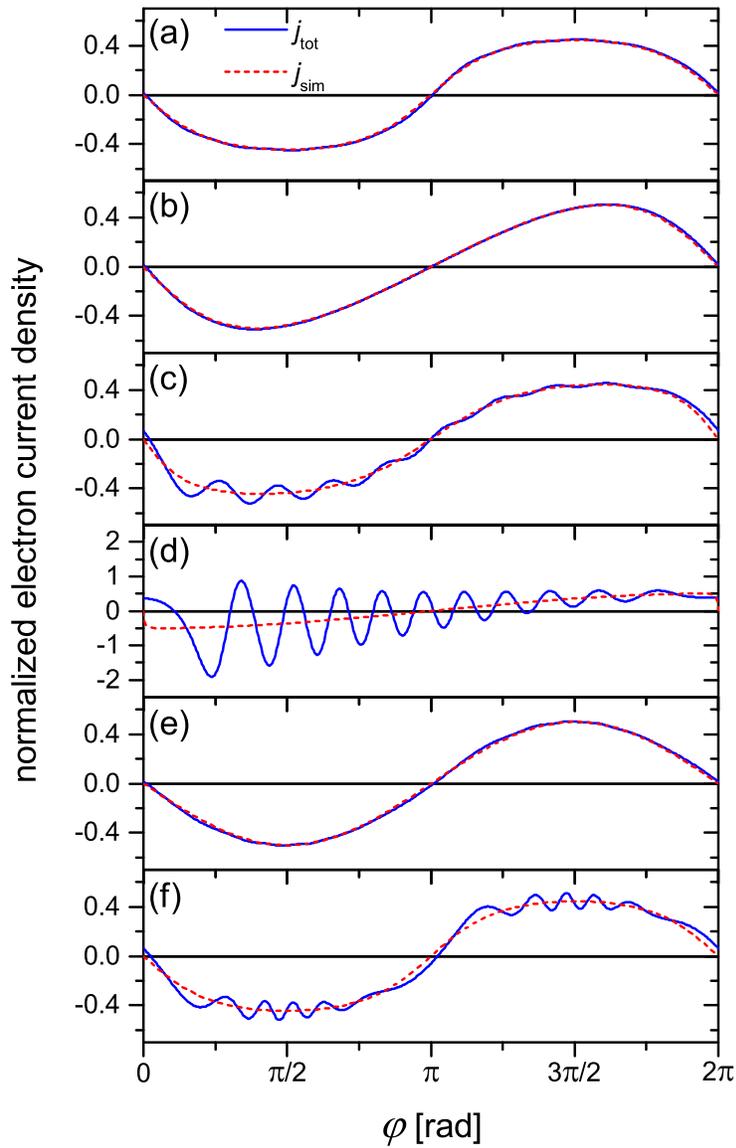}
\caption{Normalized electron current density obtained by solving the model equation (\ref{PSRequation}) numerically. For each case the solution including ($j_{tot}$, blue solid line) and neglecting ($j_{sim}$, red dashed line) the voltage drop across the plasma bulk are depicted. A single-frequency voltage waveform is applied. The other parameters are provided in table \ref{table1}.} 
\label{fig2}
\end{figure}
\begin{table}
\caption{Values of the model parameters used in the solutions of equation (\ref{PSRequation}) shown in figure \ref{fig2}. $\beta(\varphi)=0.1[1+0.5 \, \textnormal{cos}(2\varphi)]$ in \ref{fig2}(e) and \ref{fig2}(f).}
\label{table1}
\begin{indented}
\item[]\begin{tabular}{@{}l*{6}{c}}
\br
Figure \hspace{-2mm} & $\varepsilon$ & $\bar{\eta}$ & $q_{tot}$ & $a,b$ & $\beta$ & $\kappa$ \\
\mr
\ref{fig2}(a) & 1.00 & 0.00  & 1.00 & 1.5 & 0.10 & 1.0 \\
\ref{fig2}(b) & 0.50 & -1/3   & 1.15 & 1.0 & 0.10 & 1.0 \\
\ref{fig2}(c) & 0.50 & -1/3   & 1.15 & 1.5 & 0.10 & 1.0 \\
\ref{fig2}(d) & 0.01 & -0.98 & 1.41 & 1.0 & 0.10 & 1.0 \\
\ref{fig2}(e) & 1.00 & 0.00  & 1.00 & 1.0 & $\beta(\varphi)$ & 1.0 \\
\ref{fig2}(f)  & 1.00 & 0.00  & 1.00 & 1.5 & $\beta(\varphi)$ & 1.0 \\
\br
\end{tabular}
\end{indented}
\end{table}
\clearpage

This ordinary, inhomogeneous, nonlinear, second order differential equation appears to be quite cumbersome. In fact, an analytical approximation for arbitrary values of $a$, $b$, $\varepsilon$, and $\beta(\varphi)$ cannot be found. However, it can be solved for $q(\varphi)$ numerically \cite{Maple}. Note that a system described by equation (\ref{PSRequation}) resembles an oscillator, which is externally driven by $\bar{\eta}+\bar{\phi}_{\sim}(\varphi)$. The restoring force is due to the inertia of the bulk electrons, while the oscillations are damped by the momentum loss of electrons in collisions with the background gas. The resonance frequency of the oscillator may vary as function of time, but it will always be well above the applied driving frequency. This means that an efficient energy coupling from the external driver to the series circuit is possible only if the oscillator is nonlinear. Therefore, the nonlinearity on the right hand side of equation (\ref{PSRequation}) is a key feature for the self-excitation of PSR oscillations. If the variation in $\beta(\varphi)$ was negligible, the oscillator would be linear for $a=1$, $b=1$, and $\varepsilon=1$, but nonlinear for all other combinations of $a$, $b$, and $\varepsilon$. %Furthermore, the oscillation is harmonic for a stationary $\beta$ only.

We exemplarily show solutions of the normalized electron current density, $\bar{j}(\varphi)=-\dot{q}$, in figure \ref{fig2} to visualize the effect of the different parameters on the PSR self-excitation. Here and everywhere below, $q_{tot}$ is determined from the symmetry parameter, $\varepsilon$, and the maximum, $\phi_{\sim}^{max}$, and minimum, $\phi_{\sim}^{min}$, of the applied voltage waveform via $q_{tot}=[(\phi_{\sim}^{max}-\phi_{\sim}^{min})/(\phi_{tot}(1+\varepsilon))]^{1/2}$, as defined in Ref. \cite{EAE2}. The normalized applied voltage waveform for a single-frequency excitation is $\bar{\phi}_{\sim}(\varphi)=\textnormal{cos}(\varphi)$. The bulk parameter $\beta$ is set as a constant, $\beta=0.1$, in figures \ref{fig2}(a)-(d) and is assumed to vary as $\beta(\varphi)=0.1[1+0.5 \, \textnormal{cos}(2\varphi)]$ in figures \ref{fig2}(e)-(f); this parametrization is chosen, because it resembles $\beta(\varphi)$ obtained from the PIC/MCC simulation. The possibility for the self-excitation of PSR oscillations can be switched on and off by including and neglecting the voltage drop across the plasma bulk, respectively. All other parameters are specified in table \ref{table1}. The parameters $a$ and $b$ are either $1.0$ (matrix sheath) or $1.5$, as we typically find $a\approx 1.5$ in fits of equation (\ref{Usp}) to normalized PIC/MCC simulations data \cite{PSR_EAE_Letter}. Here, we assume $a$ and $b$ to be equal. This is justified, because the difference between $a$ and $b$ found in our simulations is less than one percent.
%, although the situation might be different in expanding or magnetized plasmas, for instance.
This study is focused on discharges that are symmetric for symmetric applied voltage waveforms, i.e. the EAE is the only reason of asymmetry. Many additional effects might occur in more complex situations such as geometrically asymmetric and magnetized plasmas. In particular, the sheath charge-voltage relations may be different at both electrodes ($a\neq b$) if an asymmetric magnetic field is applied, i.e. if the magnetic field strength is different at both electrodes \cite{Trieschmann}. Such situations with an additional asymmetry would further enhance the role of the PSR for the electron heating in electrically asymmetric discharges, but are far beyond the scope of this work.

The outcome of this parameter variation can be summarized as follows: (i) For a symmetric discharge ($\varepsilon=1$), the self-excitation of PSR oscillations will not occur for a constant $\beta$ (figure \ref{fig2}(a)), but the shape of the unperturbed current waveform changes as a function of $a$ and $b$ (see figures \ref{fig2}(e),(f)). (ii) An intermediate asymmetry of $\varepsilon=0.5$ is not a sufficient condition for the self-excitation of PSR oscillations under the assumption of a quadratic sheath charge-voltage relation (figure \ref{fig2}(b)); PSR oscillations can be observed, however, if the sheath charge-voltage relation exhibits a cubic component (figure \ref{fig2}(c)). This is because the sum $\bar{\phi}_{sp}+\bar{\phi}_{sg}$ is a linear function for $\varepsilon=1$ and $a=b=1$, but it becomes nonlinear when the cubic component in the sheath charge-voltage relations is taken into account. This nonlinearity is important for the correct modelling of the current waveform and the self-excitation of PSR oscillations, especially for weak and intermediate asymmetries. (iii) Certainly, the excitation strength of the PSR and, hence, the amplitude of PSR oscillations in the current are much larger for strongly asymmetric discharges ($\varepsilon=10^{-2}$, see figure \ref{fig2}(d)). Then, the sum $\bar{\phi}_{sp}+\bar{\phi}_{sg}$ is nonlinear for all choices of $a$ and $b$. Also, note that the unperturbed current waveform shows a steep gradient at $\varphi=0$. Hence, the PSR is self-excited when the absolute value of $\ddot{q}$ is largest, as this corresponds to the time, when the potential energy of the oscillator is largest. (iv) The PSR will be self-excited in symmetric discharges, if the bulk parameter $\beta$ exhibits a dependence on the RF phase and if the sheath charge-voltage relations deviate from a simple quadratic behavior. However, such a modulation alone does not cause an excitation of the PSR (figure \ref{fig2}(e)). In addition, as mentioned above the nonlinearity of $\bar{\phi}_{sp}+\bar{\phi}_{sg}$ induced by the cubic component of the sheath charge-voltage relation is a necessary condition. The combination of these two effects leads to the self-excitation of PSR oscillations even in symmetric CCRF plasmas (see figure \ref{fig2}(f)) \cite{PSR_EAE_Letter}.

In the comparison of the electron conduction current obtained from this model and PIC/MCC simulations, we will use the $\varepsilon$, $\bar{\eta}$, and $\beta$ parameters for the model as obtained from the simulations, and $a,b$ from fits of the normalized sheath charge-voltage relations to the simulation results. Furthermore, the normalized collision frequency is estimated to be $\kappa \approx 2.4$. The corresponding collision frequency of $\nu_m \approx 2.0 \times 10^8$ s$^{-1}$ is relatively high given that the argon pressure is 3 Pa, but this value is conform with the recent findings of \textit{Lafleur et al.} \cite{Lafleur_collision} under similar conditions and the good agreement of the resulting damping with the attenuation of the PSR oscillations observed in the simulations. The amplitude of the non-normalized current density is determined in the model by multiplication of the normalized current density, $\bar{j}(\varphi)$, with the factor $e \, \omega \int_0^{s_{max}} n_i(z) dz$, which is obtained from the simulations. 

In the model analysis of the electron heating, we restrict ourselves to relative values. Generally, the dissipated electric power is proportional to $j^2$. The model provides the great opportunity that we can distinguish between electron heating caused by the PSR oscillations (so-called Nonlinear Electron Resonance Heating, NERH), $P_{PSR}$, and the electron heating without PSR (by neglecting any current perturbations), $P_{sim}$:
\begin{eqnarray}
\label{Psim}
\int_0^\varphi P_{sim}(\varphi') d\varphi' & \propto & \int_0^\varphi j_{sim}^2(\varphi') d\varphi', \\
\label{Ptot}
\int_0^\varphi P_{tot}(\varphi') d\varphi' & \propto & \int_0^\varphi j_{tot}^2(\varphi') d\varphi', \\
\int_0^\varphi P_{PSR}(\varphi') d\varphi' & \propto & \int_0^\varphi \left[ j_{tot}^2(\varphi')-j_{sim}^2(\varphi') \right] d\varphi'.
\label{Ppsr}
\end{eqnarray}
The total current density, $j_{tot}=j_{sim}+j_{PSR}$, and the simplified current density, $j_{sim}$, are obtained by switching the bulk term in the voltage balance (equation (\ref{PSRequation})) on and off. Naturally, $P_{sim}+P_{PSR}=P_{tot}$, as the definition of $P_{PSR}$ in equation (\ref{Ppsr})  includes both components of $j_{tot}^2-j_{sim}^2=j_{PSR}^2+2j_{sim}j_{PSR}$, because both components depend on the PSR current density, $j_{PSR}$. 

As will be shown below, the electron heating adjacent to either electrode occurs mainly during the expansion phases of the respective sheath. The expansion of the sheath region adjacent to the powered electrode is associated with an increase of the charge in the same sheath, i.e. $\dot{q} \geq 0$ and with a negative electron current, while the expansion of the grounded electrode sheath is related to a positive electron current. Thus, the electron power absorption can be divided into the two regions of the powered electrode half space ($0 \leq z \leq d/2$) and the grounded electrode half space ($d/2 \leq z \leq d$) by making use of the current sign:
\begin{eqnarray}
\label{Ppow}
\int_0^\varphi P_{p}(\varphi') d\varphi' &=&  \int_0^\varphi P(\varphi')\Theta\left[ -j(\varphi') \right] d\varphi', \\
\int_0^\varphi P_{g}(\varphi') d\varphi' &=&  \int_0^\varphi P(\varphi')\Theta\left[ j(\varphi') \right] d\varphi'.
\label{Pgrnd}
\end{eqnarray}
These definitions are applicable to the individual components of the electron heating, $P_{PSR}$ and $P_{sim}$, as well as to the total electron heating, $P_{tot}$, by simply substituting $P$ and $j$ from equations (\ref{Psim})-(\ref{Ppsr}) into equations (\ref{Ppow}) and (\ref{Pgrnd}), respectively. Such an analysis will be most useful to investigate the role of the PSR self-excitation on the symmetry of the electron heating in CCRF plasmas.

\section{Results}
\subsection{Self-excitation of PSR oscillations in multi-frequency CCRF plasmas}
\label{Results1}

\begin{figure}[tbp]
\centering
\includegraphics[width=0.999\textwidth]{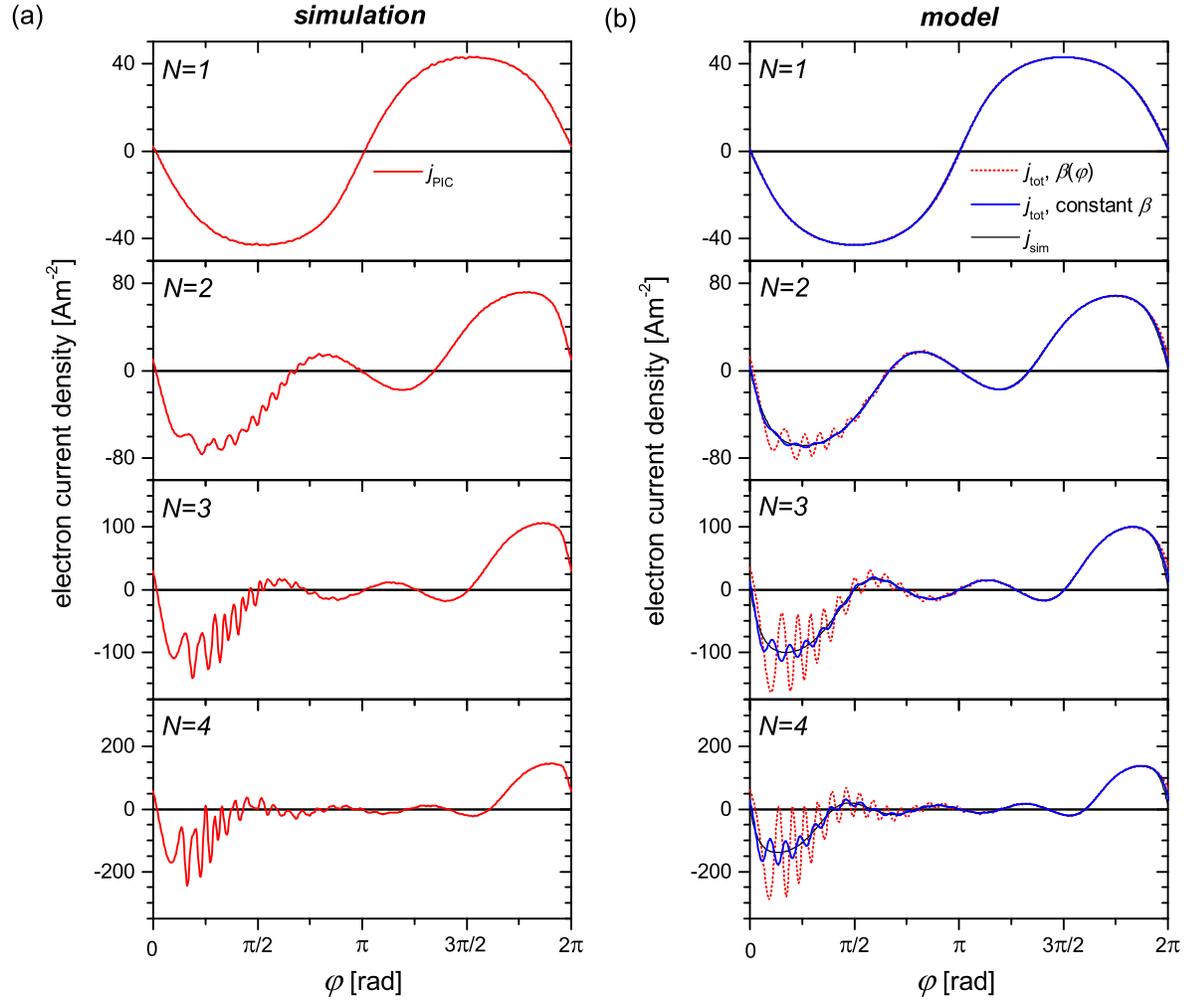}
\caption{Electron current density within one RF period for different numbers of applied harmonics, $N$, resulting from (a) PIC/MCC simulations and (b) the model. The model curves are obtained with all parameters taken from the simulations (red dotted lines), using a temporally averaged value for $\beta$ (blue solid lines) as well as the unperturbed current waveform without bulk voltage (black thin solid lines). The conditions in the PIC/MCC simulations are: Ar, 3 Pa, $d=30$ mm, $\phi_{tot}=800$ V, $\theta_k=0 \, \forall \, k$.} 
\label{fig3}
\end{figure}

Figure \ref{fig3} shows the electron current density obtained from PIC/MCC simulations and the model for different numbers of applied harmonics. Here, all phase shifts are set to zero (see figure \ref{fig1}(a)). All the input parameters of the model are deduced from the PIC/MCC simulation results. We find good agreement in the general shape of the current waveform and the current density amplitude resulting from the simulations and the model if the bulk voltage is modelled realistically, i.e. using a time dependent $\beta$. For $N=1$, the current density remains approximately sinusoidal, as one might expect in this symmetric case \cite{UCZ_sheath_model,PSR_EAE_Letter}. With increasing number of harmonics the unperturbed current density exhibits multiple local maxima and minima, since it contains all of the applied frequencies. Furthermore, the global maximum and minimum of the current density obtained from the model without PSR, i.e. neglecting the voltage drop across the plasma bulk, have the same absolute value. This symmetry is due to the phase shift of about $\pi/2$ between (all the harmonics of) the current and the applied voltage \cite{EAE12,EAE13,EAE15,EAE16,PSR_EAE_geomasymm,ELIAS1}. 

The self-excitation of PSR oscillations can be observed in all multi-frequency cases ($N \geq 2$ in figure \ref{fig3}), whereas the amplitude of these high-frequency oscillations becomes larger as a function of $N$. Moreover, the excitation strength found in the PIC/MCC simulations can only be reproduced in the model, if the temporal variation of the bulk parameter $\beta(\varphi)$ is taken into account; using the RF period averaged value $(2\pi)^{-1} \int_0^{2\pi} \beta(\varphi) d\varphi$ leads to a strong underestimation of the PSR oscillation amplitude. Thus, it can be concluded that the correct treatment of the bulk inductance (as well as of the sheath charge-voltage relation) is of critical importance for a realistic modelling of the PSR self-excitation. The strongly simplifying assumptions of a constant bulk parameter and a quadratic sheath charge-voltage relation can be regarded as main reasons for the discrepancy between the current density obtained from theoretical models and simulations/experiments in previous works \cite{PSR_Klick,PSR_diagnostics_Schulze,PSR_beams_Schulze,PSR_Ziegler1,PSR_stoch_Schulze,EAE13,PSR_Bora3,PSR_EAE_geomasymm}.

\begin{figure}[tbp]
\centering
\includegraphics[width=0.55\textwidth]{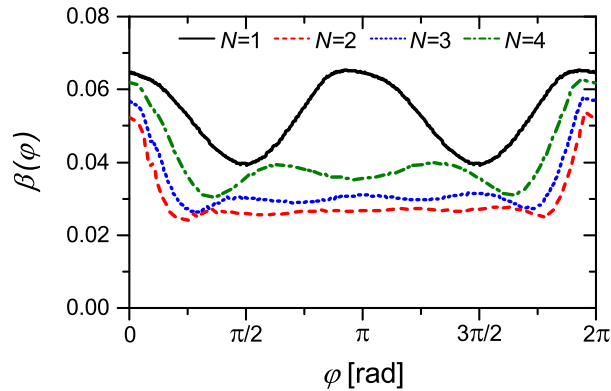}
\caption{Bulk parameter $\beta(\varphi)$ within one RF period for $N=$ 1 (black solid line), $N=$ 2 (green dash-dotted line), $N=$ 3 (blue dotted line), and $N=$ 4 (red dashed line) resulting from PIC/MCC simulations. Discharge conditions: Ar, 3 Pa, $d=30$ mm, $\phi_{tot}=800$ V, $\theta_k=0 \, \forall \, k$.} 
\label{fig1a}
\end{figure}

The temporal dependence of the bulk parameter $\beta(\varphi)$ is depicted in figure \ref{fig1a}. It typically changes by almost a factor of two under the conditions investigated here. It is maximum at the times of one of the two sheaths being fully collapsed. This is due to the fact that the effective electron plasma frequency, $\bar{\omega}_{pe}$, is smallest at those times, because the electron density approximately equals the ion density in the bulk region and is depleted in the sheath regions. If one of the sheaths is fully collapsed, the density of the electrons follows the decrease of the density of the ions towards the electrode. Therefore, the inductance of the plasma bulk is effectively enhanced due to the locally reduced electron density. In the multi-frequency cases, the density profiles are asymmetric and $\beta(\varphi)$ is maximum when the powered electrode sheath collapses, i.e. around $\varphi=0$. Under these conditions, the mean ion energy is higher at the powered electrode compared to the grounded electrode, so that the decrease in the ion density is stronger in the powered electrode sheath \cite{EAE3}. Thus, the smallest electron density as a function of space and time in the plasma bulk region is found at the powered electrode at the time of the local sheath collapse.

\begin{figure}[tbp]
\centering
\includegraphics[width=0.65\textwidth]{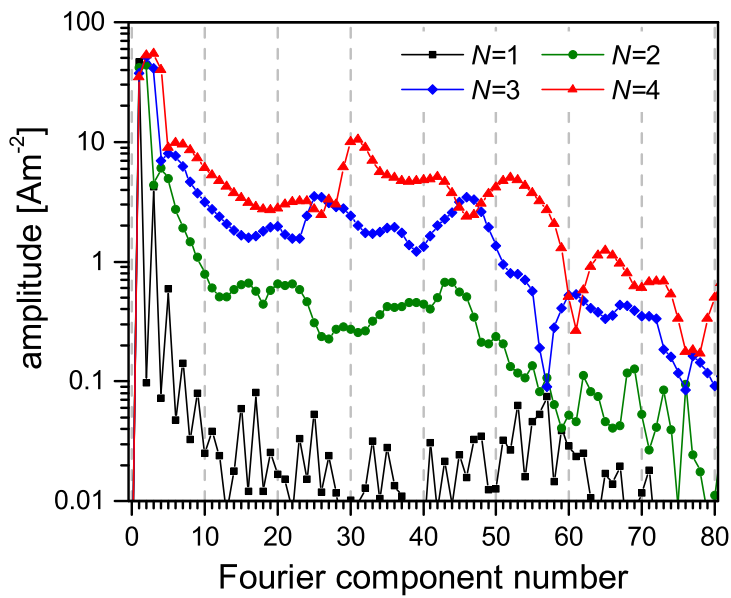}
\caption{Fourier spectrum of the electron current density obtained from PIC/MCC simulations at $N=1$ (black squares), $N=2$ (green circles), $N=3$ (blue diamonds), and $N=4$ (red triangles), respectively. Discharge conditions: Ar, 3 Pa, $d=30$ mm, $\phi_{tot}=800$ V, $\theta_k=0 \, \forall \, k$.} 
\label{fig4}
\end{figure}

A more detailed analysis of the current density via Fourier transform, shown in figure \ref{fig4}, reveals that the PSR oscillations lead to a broad frequency spectrum \cite{PSR_Mussenbrock,PSR_Ziegler1,PSR_Lieberman1,PSR_Lieberman2,PSR_ICP_Kempkes,PSR_Wang,PSR_Bora3,PSR_Bora1,PSR_Bora2}. Naturally, the amplitude is highest at the Fourier components of the applied frequencies. In the single frequency discharge, only odd Fourier components are allowed due to the mandatory symmetry of the current \cite{sheath_nonlinearity_Klick}. The amplitude quickly drops for larger frequencies. By applying an electrically asymmetric voltage waveform ($N \geq 2$), this symmetry restriction is removed and all higher harmonics of the fundamental frequency may occur in the discharge current. The main PSR frequencies, which also dominate the current perturbations depicted in figure \ref{fig3}, correspond to the peak around the 25th ($N=$ 3) and 30th ($N=$ 4) harmonic of the fundamental applied voltage frequency (see figure \ref{fig4}). These frequencies are in the 311 MHz - 407 MHz range. The strength of the PSR self-excitation, i.e. the amplitude of the current branch at higher frequencies, increases as a function of $N$. This is primarily due to two reasons: firstly, the discharge becomes more and more asymmetric ($\varepsilon\approx$ 1.00, 0.68, 0.57, and 0.52 at $N=$ 1, 2, 3, and 4) due to the enhanced asymmetry of the plasma density profile and the self-amplification of the EAE at low pressures \cite{PSR_EAE_Zoltan,EAE2,EAE3,EAEmultif1,EAEmultif2}. Secondly, the bulk parameter $\beta(\varphi)$ changes faster, because the time interval between the collapsing and the expanding phase of the powered electrode sheath becomes shorter (see figure \ref{fig1}). As a consequence, figure \ref{fig4} shows that the PSR spectrum broadens as a function of $N$. 

\begin{figure}[tbp]
\centering
\includegraphics[width=0.999\textwidth]{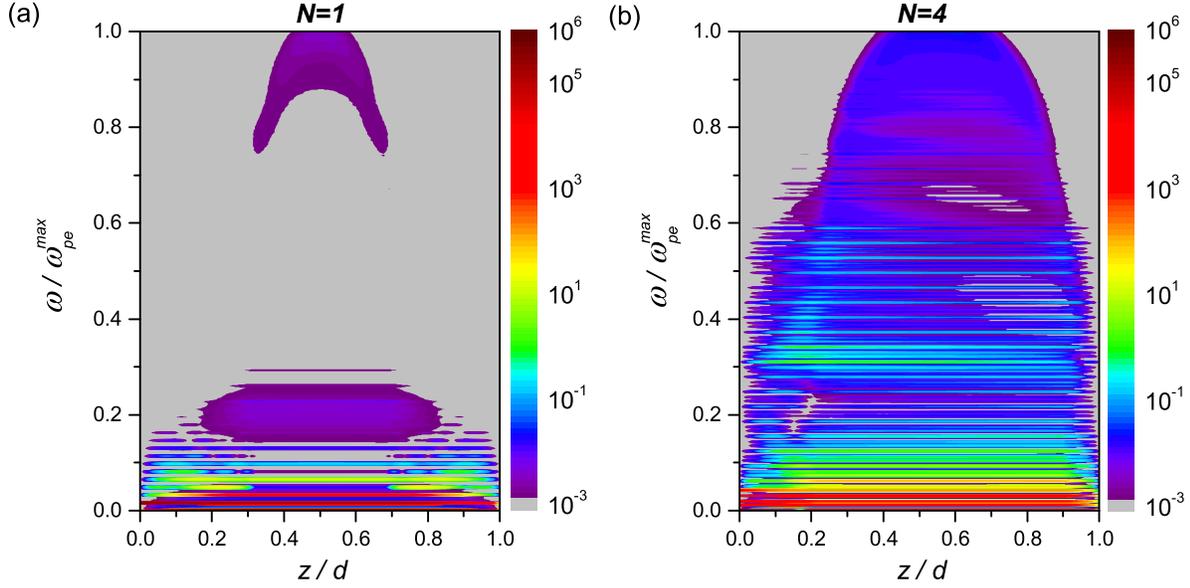}
\caption{Power spectral density of the potential, $\hat{\phi}$ (see equation (\ref{EqPSD})), as a function of position obtained from PIC/MCC simulations operated at (a) 13.56 MHz ($N=$ 1) and (b) four consecutive harmonics ($N=$ 4). The color scales give values in V$^2$. The Fourier spectrum is normalized by the maximum electron plasma frequency in each case. Discharge conditions: Ar, 3 Pa, $d=30$ mm, $\phi_{tot}=800$ V, $\theta_k=0 \, \forall \, k$.} 
\label{fig5}
\end{figure}

\begin{figure}[tbp]
\centering
\includegraphics[width=0.65\textwidth]{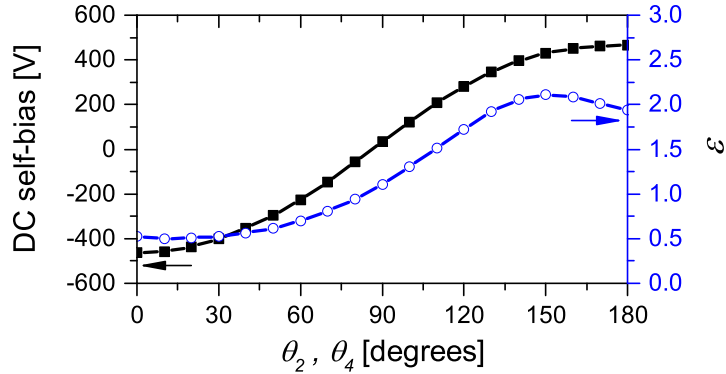}
\caption{DC self-bias (black squares) and symmetry parameter (blue circles) as a function of the phase shifts of the second and fourth harmonic, $\theta_2$ and $\theta_4$, obtained from PIC/MCC simulations operated at $N=$ 4. The phase of the fundamental frequency and the third harmonic are kept zero, $\theta_1=0$ and $\theta_3=0$. Discharge conditions: Ar, 3 Pa, $d=30$ mm, $\phi_{tot}=800$ V.} 
\label{fig7}
\end{figure}

The effect of the self-excitation of PSR oscillations on the electric potential are visualized in figure \ref{fig5}, which shows the power spectral density as a function of position between the powered ($z=0$) and grounded ($z=d$) electrodes. The spectrum is normalized by the maximum electron plasma frequency, $\omega_{pe}^{max}$, obtained from the spatio-temporally resolved electron density. The spectra consist of lines, which appear at discrete frequencies, i.e. at the applied frequencies and their higher harmonics. In the single-frequency case (figure \ref{fig5}(a)), no PSR oscillations are observed (see figure \ref{fig3}). Hence, the potential oscillates primarily at the applied frequency of 13.56 MHz ($\omega / \omega_{pe}^{max} \approx 8 \times 10^{-3}$) and about 10 of its harmonics, in the spectral region $\omega \leq 0.1 \, \omega_{pe}^{max}$. A small contribution is found around the local electron plasma frequency, i.e. at $\omega \approx \omega_{pe}^{max}$ in the discharge center. This feature follows the decrease of the electron density and of the electron plasma frequency towards the sheath regions: it originates from small local oscillations of the electron ensemble.
An asymmetry is induced in the multi-frequency case and, hence, the power spectral density of the potential exhibits a spatial asymmetry (see figure \ref{fig5}(b)). A much richer distribution is found at low frequencies, compared to the single-frequency case. This is a direct effect of the multi-frequency driving voltage waveform, generating a higher power spectral density at the consecutive harmonics of the fundamental frequency of the applied voltage waveform (between $\omega / \omega_{pe}^{max} \approx 5 \times 10^{-3}$ corresponding to 13.56 MHz and $\omega / \omega_{pe}^{max} \approx 2 \times 10^{-2}$ corresponding to 54.24 MHz). The PSR oscillations are associated with oscillations of the potential, leading to a branch in the spectral region $\omega \geq 0.3 \, \omega_{pe}^{max}$. These oscillations are stronger at the powered electrode sheath region, because this sheath is large and expands quickly at the time of strong resonance oscillations, while the grounded sheath is small.
%expanding at the time of strong resonance oscillations, and are more and more spatially constricted with increasing frequency. 
Eventually, this enhances the power density at high frequencies, i.e. up to the local electron plasma frequency. Such effects of the PSR oscillations on the local plasma properties cannot be captured in the frame of a global model, i.e. figure \ref{fig5}(b) reveals the limitations of the existing model approaches. These limitations might be overcome by a kinetic description of the self-excited PSR in the future.

\begin{figure}[tbp]
\centering
\includegraphics[width=0.65\textwidth]{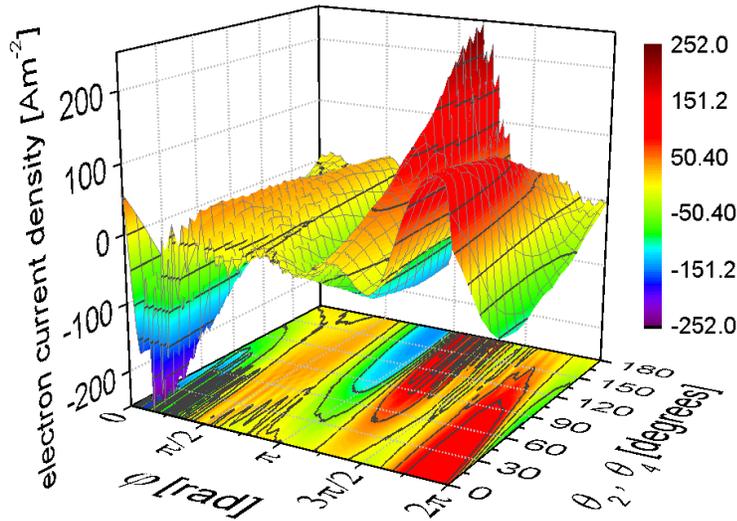}
\caption{Electron current density as a function of the RF phase, $\varphi$, and the phase shifts of the second and fourth harmonic, $\theta_2 = \theta_4$, obtained from PIC/MCC simulations operated at $N=$ 4. The phase of the fundamental frequency and the third harmonic are kept zero, $\theta_1= \theta_3=0$. Discharge conditions: Ar, 3 Pa, $d=30$ mm, $\phi_{tot}=800$ V.} 
\label{fig6}
\end{figure}

The asymmetry of the applied voltage waveform (see figure \ref{fig1}) and, hence, the asymmetry of the discharge can be controlled by tuning the phase shifts between the applied voltage harmonics. The minimum negative and maximum positive DC self-bias values are achieved by choosing either all phase shifts to be zero or phase shifts of $0^\circ$ for the odd harmonics and $180^\circ$ for the even harmonics, respectively (see figure \ref{fig7}) \cite{EAEmultif1}. As mentioned above, the model predicts that the PSR is self-excited in symmetric discharges ($\varepsilon=1$), only if the plasma bulk inductance varies temporally and the charge-voltage relation of the plasma sheaths exhibits a cubic component. In fact, we find that these two conditions are fulfilled for all phase shifts in the simulations, so that PSR oscillations are self-excited in the current density for all phase shifts (see figure \ref{fig6}) and, accordingly, for all values of the symmetry parameter that changes from $\varepsilon \approx 1/2$ at $\theta_2=\theta_4=0^\circ$ over $\varepsilon \approx 1$ at $\theta_2=\theta_4=90^\circ$ to $\varepsilon \approx 2$ at $\theta_2=\theta_4=180^\circ$ (see figure \ref{fig7}). The excitation strength and, hence, the PSR oscillations amplitude is smaller for phases around $90^\circ$, i.e. for less asymmetric discharges, though. Moreover, figure \ref{fig6} shows that the starting phase of the PSR changes from $\varphi \approx 0$ at $\theta_2=\theta_4=0^\circ$ to $\varphi \approx \pi$ at $\theta_2=\theta_4=180^\circ$, because it is excited at the time of sheath collapse adjacent to the powered and grounded electrode for $0^\circ$ and $180^\circ$ (see figure \ref{fig1}), respectively. Accordingly, it is excited more weakly, but twice in the RF period for intermediate phase shifts, such as $\theta_2=\theta_4=90^\circ$.

\subsection{Role of the PSR for the electron heating in multi-frequency CCRF plasmas}

\begin{figure}[tbp]
\centering
\includegraphics[width=0.65\textwidth]{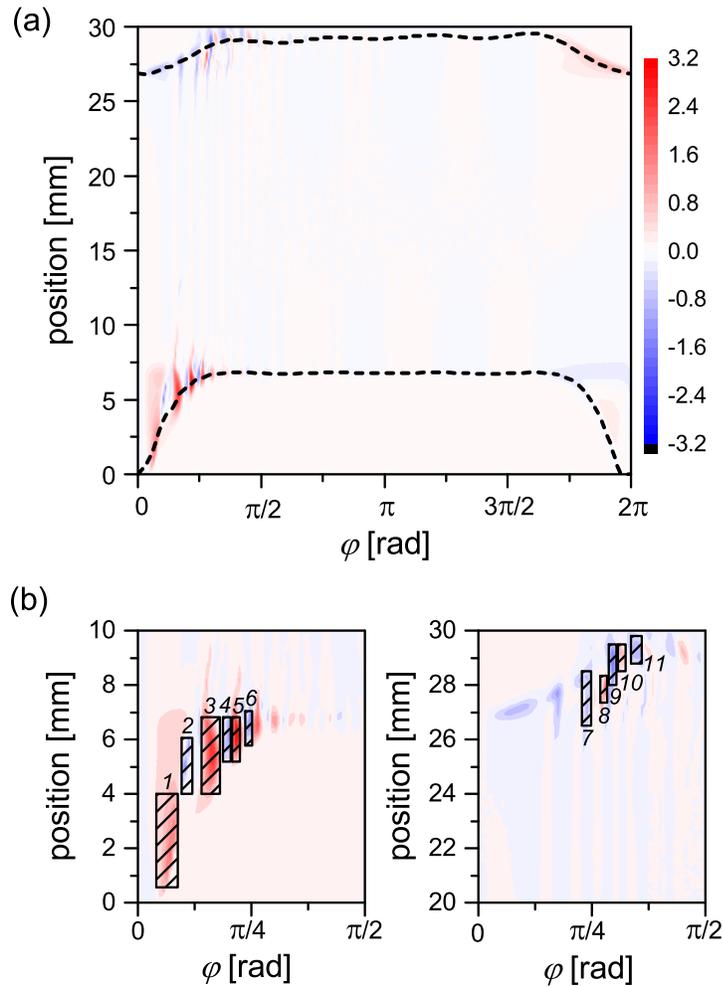}
\caption{(a) Spatio-temporal plot of the electron heating rate within one RF period between the powered electrode at 0 mm and the grounded electrode at 30 mm obtained from the PIC/MCC simulation of a discharge operated with a multi-frequency voltage waveform. Here, $N=$ 4 and all phase shifts are set to zero. The colour scale provides the heating rate in 10$^6$ Wm$^{-3}$. The dashed lines indicate the positions of the plasma sheath edges. (b) Zoom into the regions close to the powered (left) and grounded (right) electrode in the initial phase of the rf period. The hatched boxes indicate the regions, in which the EEDFs are analyzed (figure \ref{figEEDF}). Discharge conditions: Ar, 3 Pa, $d=30$ mm, $\phi_{tot}=800$ V, $\theta_k=0 \, \forall \, k$.} 
\label{fig8}
\end{figure}

\begin{figure}[tbp]
\centering
\includegraphics[width=0.65\textwidth]{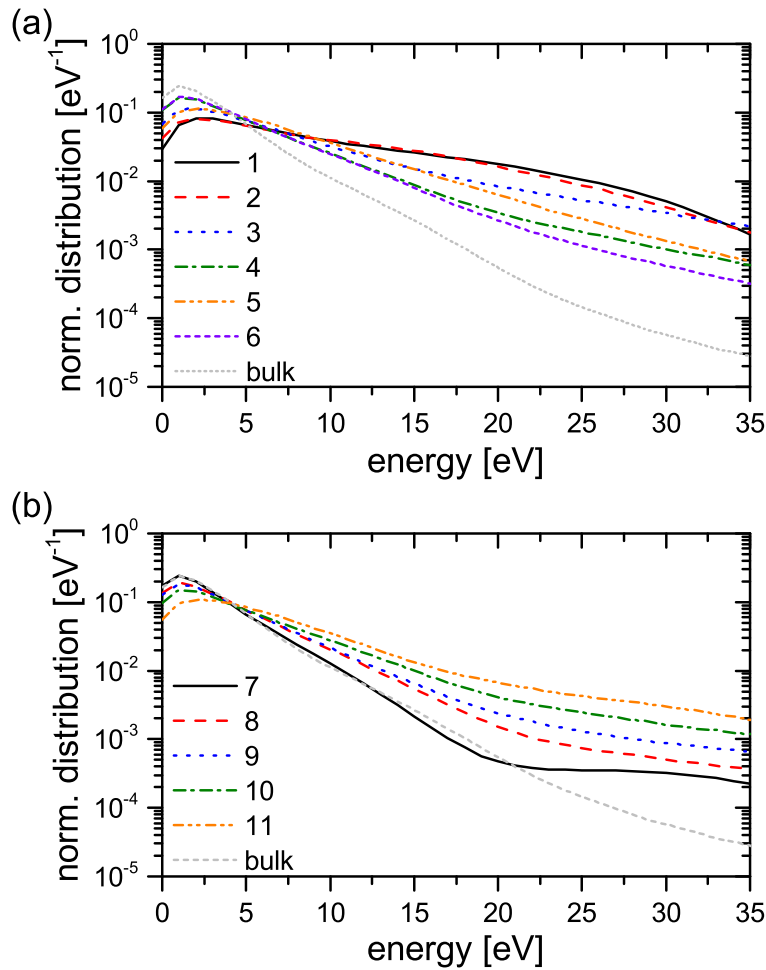}
\caption{Electron Energy Distribution Functions (EEDFs) in the regions defined in figure \ref{fig8}(b). Each EEDF is normalized by its integral, i.e. by the density of electrons in the respective region. Discharge conditions: Ar, 3 Pa, $d=30$ mm, $N=4$, $\phi_{tot}=800$ V, $\theta_k=0 \, \forall \, k$.} 
\label{figEEDF}
\end{figure}

The electron heating dynamics exhibits a great complexity in low pressure CCRF plasmas \cite{Belenguer,Fermi,Surendra,Gozadinos,Salabas,Vender,Schulze_dual,Tochikubo,Schulze_eheat,Dittmann,Kuellig,Killer,Gans,Mahony,Schulze_field_reversal,UCZ_field_reversal,ELIAS1,Perrin,Boeuf,BulkMode,BulkMode3,Goedheer,PSR_beams_Schulze,Schulze_eheat_IEEE,ambipolar}. The effects of resonances on the electron heating are not fully understood. Figure \ref{fig8} shows the spatio-temporal distribution of the electron heating rate obtained from our PIC/MCC simulation of a discharge driven by four consecutive harmonics. Apparently, the discharge is operated in the $\alpha$-mode, as the electron heating is strongest at the phases of sheath expansion in the regions around the momentary plasma sheath edge, extending to the position of the maximum sheath extension ($z\approx$ 7 mm). The heating rate of secondary electrons inside the sheaths is small due to their small number and rare multiplication by ionization collisions in the sheaths. The strongest electron heating occurs at $0 \leq \varphi\ \leq \pi/4$, close to the powered electrode. Here, the plasma series resonance is self-excited and leads to a modulation of the sheath expansion velocity. This, in turn, causes the generation of multiple energetic electron beams, which propagate towards the plasma bulk \cite{Vender,Schulze_dual,Schulze_eheat,Dittmann,Kuellig,Killer,Gans,Mahony,Schulze_field_reversal,UCZ_field_reversal,PSR_beams_Schulze,Schulze_eheat_IEEE,ambipolar}. In addition, these beams gain energy when crossing the layer of ambipolar electric field around the position of maximum sheath extension \cite{ambipolar}. Note that the electron heating rate is negative, i.e. electron cooling occurs, at the times between two rapid partial sheath expansions. At the opposing grounded sheath, there is a pattern of dominantly electron cooling, which is modulated due to the presence of PSR oscillations, as well. The electron heating by the expansion of the grounded sheath at $7 \pi / 4 \leq \varphi \leq 2 \pi$ is much smaller compared to the one at the powered electrode sheath expansion, because the maximum sheath width and the sheath expansion velocity are smaller and there is no excitation of the series resonance. The spatio-temporal distribution of the electron heating depicted in figure \ref{fig8} motivates the model approach of associating the electron heating with the expansion of the powered and grounded electrode sheath and, hence, with the sign of the electron current.

Figure \ref{figEEDF} shows the Electron Energy Distribution Functions (EEDFs) in the regions defined in figure \ref{fig8}(b). The bulk EEDF is obtained in the discharge center (14.99 mm $\leq z \leq$ 15.01 mm) time averaged over all rf phases. Each EEDF is normalized by its integral, i.e. by the density of electrons in the respective region. In general, the energy of electrons increases as the regions approach the electrodes. This is due to the fact that only the energetic fraction of the bulk electrons may move across the electric field, which is also present outside of the sheath region \cite{ambipolar}. The heating and cooling rates are relatively small in regions 1 and 2. The EEDFs in regions 1 and 2 are very similar, because the electrons that are heated in region 1 propagate towards the bulk and cross region 2 (compare figure \ref{fig11}(a)). In the regions 3 to 6, the fast change between heating and cooling due to the PSR leads to an alternating gain and loss of energy in the electron distribution function. The mean electron energy is 8.5 eV, 5.6 eV, 7.2 eV, and 5.1 eV in regions 3, 4, 5, and 6, respectively. These values are much higher than the mean energy of 3.2 eV of the electrons in the bulk EEDF. Such changes in the electron energy distribution are important for applications of CCRF plasmas, where collision processes of energetic electrons drive the plasma chemistry. Thus, the rapid changes of the EEDF due to the PSR might significantly enhance process rates in reactive gases. At the grounded electrode, the heating and cooling rates are relatively small. Therefore, the electron energy monotonically increases towards the electrode. The large population of the high energy tail of the EEDF in region 9 is caused by the arrival of beam electrons, which have been accelerated by the expansion of the opposing sheath adjacent to the powered electrode. These electrons can propagate through the entire plasma bulk at low pressures and bounce multiple times between the two sheaths \cite{bounce_res}.

\begin{figure}[tbp]
\centering
\includegraphics[width=0.65\textwidth]{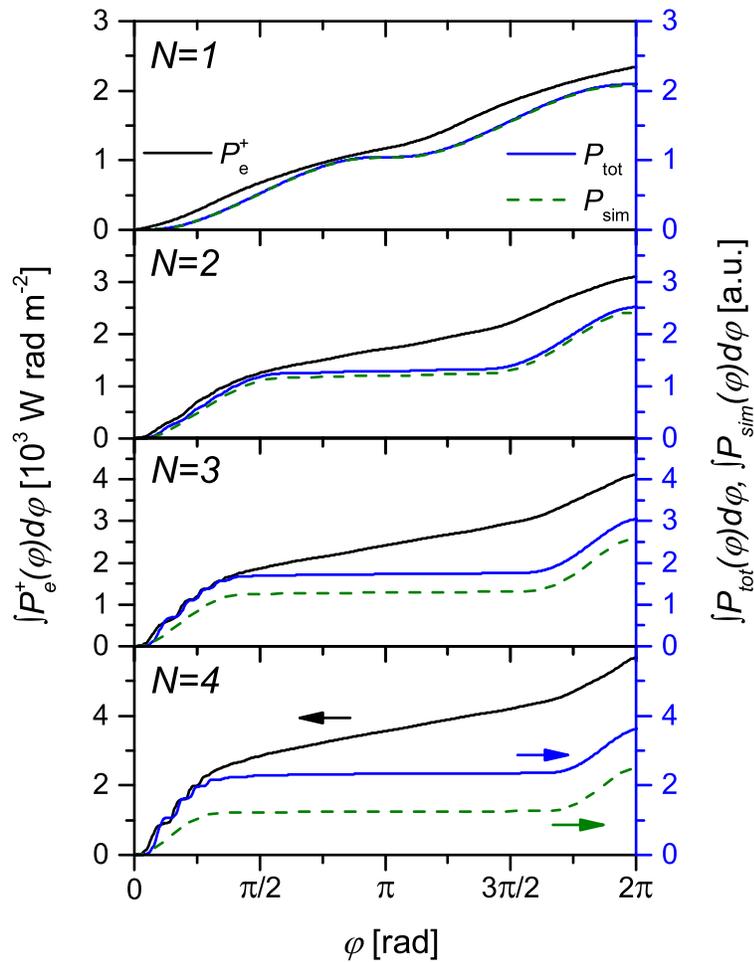}
\caption{Accumulated electron heating within one RF period for different numbers of applied harmonics, $N$, obtained from PIC/MCC simulations ($P_e^{+}$, black line, left scale) and the model including ($P_{tot}$, blue line, right scale) or excluding ($P_{sim}$, green dashed line, right scale) the PSR self-excitation. The conditions in the PIC/MCC simulation are: Ar, 3 Pa, $d=30$ mm, $\phi_{tot}=800$ V, $\theta_k=0 \, \forall \, k$.} 
\label{fig9}
\end{figure}

\begin{figure}[tbp]
\centering
\includegraphics[width=0.999\textwidth]{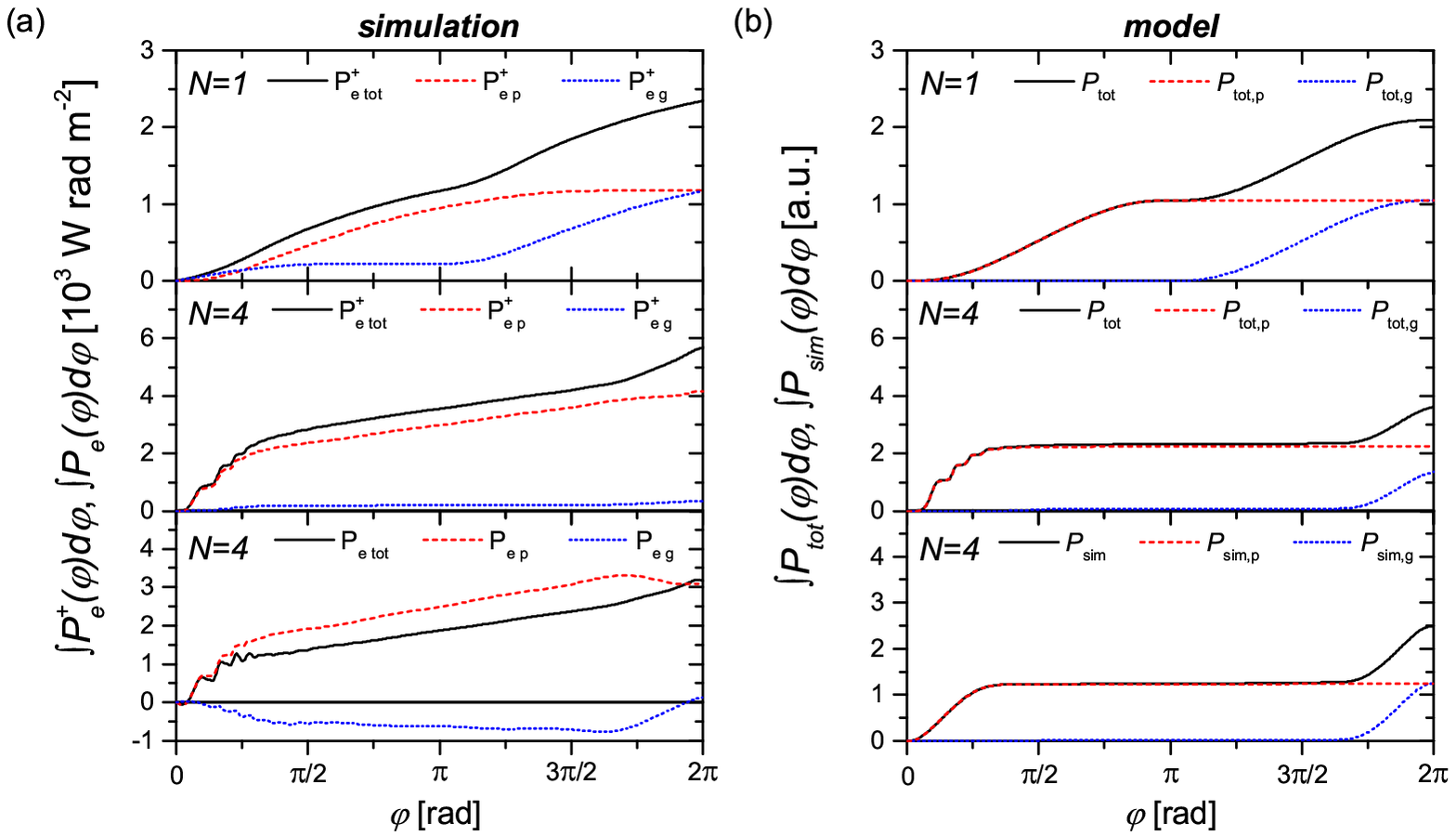}
\caption{Accumulated electron heating within one RF period obtained from (a) PIC/MCC simulations and (b) the model. The heating  in the entire discharge volume (black line) is split into the components in the two halves adjacent to the powered ($0 \leq z \leq d/2$, red line) and grounded ($d/2 \leq z \leq d$, blue line) electrodes. The discharge is driven by a single-frequency voltage waveform ($N=$ 1) in the top plots and a four-frequency voltage waveform ($N=$ 4) in the center and bottom plots, respectively. The bottom plots show the accumulated electron heating (a) including cooling obtained from the simulation and (b) excluding the PSR self-excitation obtained from the model. The conditions in the PIC/MCC simulation are: Ar, 3 Pa, $d=30$ mm, $\phi_{tot}=800$ V, $\theta_k=0 \, \forall \, k$.} 
\label{fig10}
\end{figure}

Figure \ref{fig9} shows the accumulated electron heating within one RF period as obtained from the PIC/MCC simulations for different numbers of applied harmonics. Here, only the positive contribution (heating) is considered and cooling (a negative heating rate) is neglected in order to allow for a comparison of the simulations with the model, which only includes heating.
%In all cases, the accumulated electron heating obtained from the simulations increases continuously at all times within the RF period. However, 
A plateau is formed in the accumulated electron heating resulting from the model, e.g. there is no significant electron heating between $\varphi \approx \pi /2$ and $\varphi \approx 3\pi /2$ at $N=$ 4. This difference between simulation and model is caused by secondary electrons, which are present in the simulations but their effect is neglected in the model. The acceleration and multiplication of secondary electrons is largest at the times of fully expanded sheaths. Therefore, there is an almost linear gain in the accumulated electron heating in the simulation, when the powered electrode sheath is large (see figure \ref{fig8}). The simulated behavior is very well reproduced by the model in the initial phase ($0 \leq \varphi \leq \pi /4$). This is the phase of large resonance oscillations, which modulate the electron heating (figure \ref{fig8}). The electron heating due to the PSR (so called Nonlinear Electron Resonance Heating \cite{PSR_Klick,PSR_Birdsall1,PSR_SEERS_Klick,PSR_Mussenbrock,PSR_Semmler,PSR_diagnostics_Schulze,PSR_Ziegler1,PSR_stoch_Schulze,PSR_Lieberman1,PSR_Lieberman2,PSR_EAE_Zoltan,PSR_Yamazawa,PSR_Ziegler2,PSR_Bora3,PSR_Bora4,PSR_Bora1}) is associated with a step-like behavior of the accumulated electron heating. The model allows for switching on and off the PSR self-excitation. The difference between the accumulated heating with ($P_{tot}$ in equation (\ref{Ptot})) and without ($P_{sim}$ in equation (\ref{Psim})) PSR becomes larger with increasing $N$, because the self-excitation of the PSR becomes stronger (see figure \ref{fig3}) and its effect on the electron heating becomes more and more important. The model suggests that about one third of the total electron heating within one RF period is directly related to the enhancement via the PSR in the plasma driven by four consecutive harmonics. Moreover, the step-like behavior at the beginning of the RF period will vanish, if the self-excitation of the PSR is excluded.

The division of the accumulated electron heating within the half spaces adjacent to the powered ($0 \leq z \leq d/2$, $P_{e,p}^{+}$ in equation (\ref{Pep})) and grounded ($d/2 \leq z \leq d$, $P_{e,g}^{+}$ in equation (\ref{Peg})) electrodes, respectively, is depicted in figure \ref{fig10}. The total electron heating in both halves is identical within one RF period in the single-frequency case, so we find a perfect symmetry in both the simulation and the model. The only difference between the two half spaces is that heating occurs 180$^\circ$ out of phase at either side due to the phase difference of the two sheaths (see figure \ref{fig1}). The picture will change completely, if the discharge is driven by a multi-frequency voltage waveform. Then, there is a strong spatial asymmetry in the accumulated electron heating. In consistency with the simulation, the model indicates that the electron heating adjacent to the powered electrode is much larger than the one adjacent to the grounded electrode. 
Small details such as the heating of electrons during the sheath collapse at the grounded electrode, which is visible in figure \ref{fig8}, are based on kinetic effects and are not fully captured by the model, although $P_{tot,g} >0$ for $\varphi \geq \pi/2$.
Within one RF period, the heating at both sides differs by a factor of almost two in the model; in the simulation this factor is additionally enlarged by the heating of secondary electrons, which is stronger at the powered electrode sheath due to the much longer period of a largely expanded sheath compared to the grounded electrode sheath. Again, the model provides the opportunity of switching on and off the self-excitation of PSR oscillations. In this way, it can be proven that the spatial asymmetry in the electron heating obtained from the model is solely due to the presence of the series resonance, because it would vanish if there were no resonance (see bottom plot of figure \ref{fig10}(b)); the asymmetry in the simulations is due to this effect and due to the difference of the electron heating of secondary electrons in the two sheaths. The step-like increase of the accumulated electron heating at the powered electrode associated with the PSR self-excitation will disappear, if the bulk term is neglected in the voltage balance (equation (\ref{PSRequation})). Then, the electron heating remains symmetric. The asymmetry is even further enhanced, when cooling is included in the simulation analysis. The bottom plot of figure \ref{fig10}(a) shows that there is almost no net heating in the discharge half adjacent to the grounded electrode. Moreover, the cooling between the phases of rapid partial expansion of the powered electrode sheath (see figure \ref{fig8}) leads to a repetitive change of the slope in the initial phase ($0 \leq \varphi \leq \pi/4$). Again, the almost linear increase of the accumulated electron heating at the powered electrode for intermediate RF phases ($\pi/2 \leq \varphi \leq 3\pi/2$) is caused by secondary electrons.

Moreover, a comparison of the bottom panels of figures \ref{fig10}(a) and (b) indicates that the self-excitation of the PSR leads to an enhancement of the net heating. After the initial phase (dominant heating close to the powered electrode), the accumulated heating and cooling in the simulation is similar to the accumulated heating obtained from the model excluding the effect of the PSR. Therefore, the net heating (i.e. the sum of the additional heating and the additional cooling) caused by the PSR self-excitation largely compensates the cooling that would be present without PSR. Thus, we conclude that there is a positive net heating due to the PSR.

\begin{figure}[tbp]
\centering
\includegraphics[width=0.65\textwidth]{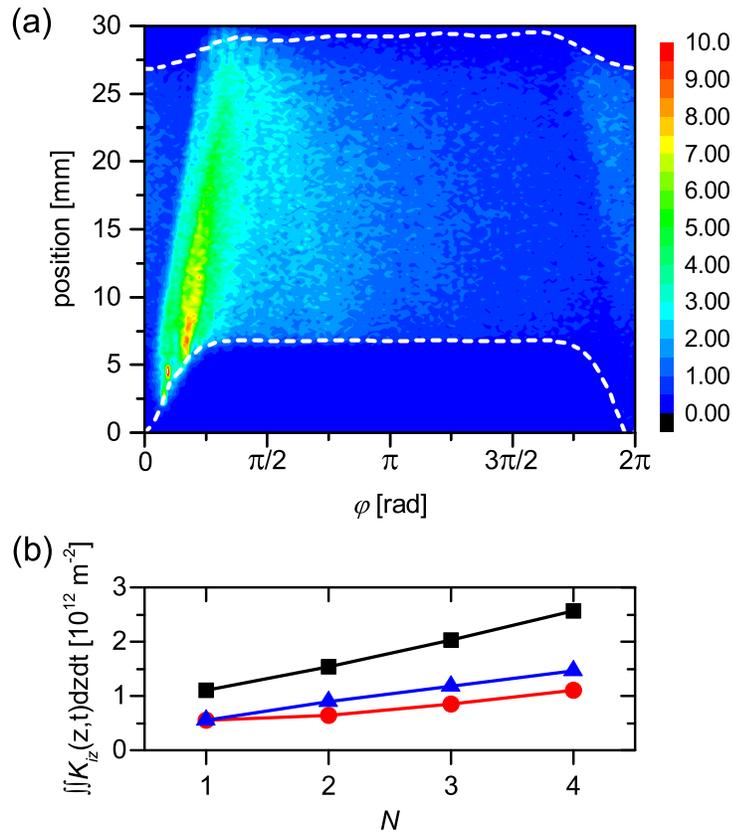}
\caption{(a) Spatio-temporal plot of the ionization rate, $K_{iz}$, within one RF period between the powered (0 mm) and grounded (30 mm) electrodes obtained from the PIC/MCC simulation of the discharge driven by a four-frequency voltage waveform ($N=$ 4). The plasma sheath edges are drawn as white dashed lines. The color scale provides the ionization rate in 10$^{21}$ m$^{-3}$ s$^{-1}$. (b) Total ionization (black squares) and ionization in the two halves adjacent to the powered ($0 \leq z \leq d/2$, red circles) and grounded ($d/2 \leq z \leq d$, blue triangles) electrodes as a function of the number of applied harmonics, $N$, obtained from PIC/MCC simulations. Discharge conditions: Ar, 3 Pa, $d=30$ mm, $\phi_{tot}=800$ V, $\theta_k=0 \, \forall \, k$.} 
\label{fig11}
\end{figure}

\begin{figure}[tbp]
\centering
\includegraphics[width=0.65\textwidth]{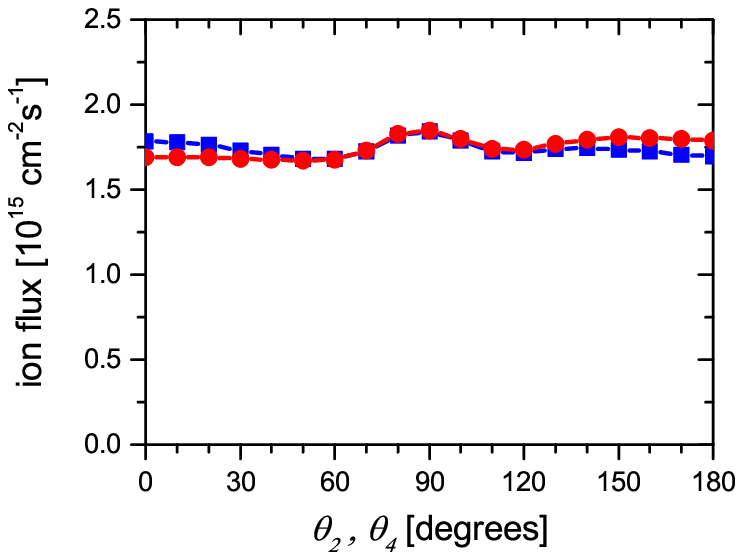}
\caption{Ion flux onto the powered (blue squares) and grounded (red circles) electrodes within one RF period as a function of the phases $\theta_2 = \theta_4$ obtained from PIC/MCC simulations operated at $N=$ 4. The phase of the fundamental frequency and the third harmonic are kept zero, $\theta_1=0$ and $\theta_3=0$. Discharge conditions: Ar, 3 Pa, $d=30$ mm, $\phi_{tot}=800$ V.}
\label{fig13}
\end{figure}

The asymmetry of the electron heating and of the discharge properties causes a spatial asymmetry in the ionization, too. Figure \ref{fig11}(a) shows the spatially and temporally resolved ionization rate obtained from a multi-frequency ($N=$ 4) PIC/MCC simulation. It can be clearly observed that the energetic electron beams, which are launched by the rapid partial expansion of the powered electrode sheath, propagate through the plasma bulk. As discussed above, the generation of multiple beams is facilitated by the modulation of the sheath expansion velocity due to the self-excitation of the PSR. A fraction of these energetic electrons reaches the opposing sheath, as the electron mean free path is comparable to the length of the plasma bulk, where it is reflected back towards the discharge center. Therefore, in this non-local, low pressure regime the dominant electron heating is strongly localized (see figure \ref{fig8}), but the subsequent ionization is distributed over a large region in space and time. In particular, the ionization by electrons, which undergo one or more reflections at the plasma sheath edges, is spatially asymmetric. This is due to the fact that the powered sheath is largely expanded and the grounded electrode sheath is small in the important interval ($\varphi \geq \pi/8$). Thus, electrons are confined in the plasma bulk region, which is closer to the grounded electrode than to the powered one. Accordingly, in the multi-frequency cases considered here the total ionization in the powered electrode half is smaller than the one in the grounded electrode half of the discharge volume, although the electron heating is much stronger in the powered electrode half (figure \ref{fig11}(b)). This non-local ionization dynamics helps maintaining an approximately constant ion flux to either electrode, independent of the phase shifts between the applied voltage harmonics (see figure \ref{fig13}). Thus, the control of the ion energy at constant ion flux via the EAE, i.e. by tuning the phase shifts, as discussed in great detail in previous works \cite{EAE2,EAE3,EAE5,ELIAS3,EAEmultif1,EAEmultif2,PSR_EAE_geomasymm}, is not much disturbed by the self-excitation of PSR oscillations for two reasons: firstly, the PSR is self-excited not only in asymmetric CCRF plasmas but PSR oscillations can be observed in the current of multi-frequency CCRF plasmas for all phase shifts and discharge asymmetries, respectively. Therefore, an enhancement of the electron heating is found for all applied multi-frequency voltage waveforms. (Note that the present situation ($N=4$) differs from a previous study of a dual-frequency ($N=2$) discharge, where the electron heating has been found to be strongly enhanced for asymmetric applied voltage waveforms only \cite{PSR_EAE_Zoltan}.) Secondly, this heating enhancement is localized, but the energy relaxation length is large and ionization by electron-neutral collisions is broadly distributed over the entire plasma bulk region.

\section{Conclusions}

Plasma series resonance oscillations are self-excited in low-pressure CCRF plasmas driven by multiple consecutive voltage harmonics. Up to now, the self-excitation of the PSR has only been observed in asymmetric capacitive discharges. Using a combined approach of PIC/MCC simulations and an equivalent circuit model, we demonstrate that the self-excitation of the PSR occurs in geometrically symmetric capacitive discharges driven by either symmetric or asymmetric voltage waveforms. The nonlinearity in the governing equation, which is required to self-excite high frequency PSR oscillations of the current, is found to be strongly affected by the temporal modulation of the bulk inductance and the cubic charge-voltage relation of the plasma sheaths. We find that both effects are important for the self-excitation of the PSR, and that the electron current density resulting from the self-consistent simulations can only be reproduced by the model if these effects are taken into account.

The Fourier analysis of the current density and the potential distribution between the two electrodes provide further insight. A broad spectrum of higher harmonics is observed in the presence of PSR oscillations and the main frequency branch of the PSR oscillation is about ten times higher than the applied voltage frequencies. The results of the kinetic simulations show that the self-excited PSR affects the local plasma properties, as it facilitates oscillations of the potential profile in the entire spectrum between the applied frequencies and the local electron plasma frequency. These oscillations are stronger adjacent to the sheath, which expands during the phase of a large PSR amplitude. The self-excitation of the PSR can be initiated at the collapse of either the powered and/or the grounded electrode sheath, depending on the shape of the applied voltage waveform and the plasma asymmetry. 

The series resonance plays an important role in the electron heating dynamics. In particular, it leads to a modulation of the sheath expansion velocity and, thereby, to rapid oscillations between heating and cooling of electrons around the region of the momentary plasma sheath edge. The electron heating is enhanced by this Nonlinear Electron Resonance Heating (NERH). Moreover, the electron heating becomes spatially asymmetric in the presence of PSR oscillations, e.g. there will be more heating in the discharge half adjacent to the powered electrode, if the PSR is initiated at the time of powered electrode sheath collapse and electrons are primarily heated around the expanding edge of the powered electrode sheath. Based on the model results it can be concluded, that the asymmetry of the electron heating is caused by the additional heating due to the series resonance oscillations, which approximately doubles the electron heating in the discharge half adjacent to the powered electrode. The asymmetry will be even stronger, if secondary electrons as well as electron cooling is included in the simulation analysis: then, there is almost no net heating in the discharge half adjacent to the grounded electrode. 

However, the spatial asymmetry of the ionization is reversed at our low pressure conditions. This is due to the electron kinetics: electrons, that have gained high energies by the NERH around the expanding edge of the powered electrode sheath, propagate through the plasma bulk in beam-like structures. The energy relaxation occurs, therefore, largely in the grounded electrode half sphere, because these beams are confined between the small grounded electrode sheath and the fully expanded powered sheath. Thus, the enhancement and asymmetry of the electron heating induced by the self-excitation of PSR oscillations does not cause a disturbance of the separate control of ion energy and ion flux at both electrodes via the Electrical Asymmetry Effect.

In the future, these findings should be verified experimentally and the role of the PSR in processing applications of multi-frequency capacitive plasmas needs to be investigated in detail. The rapid change of the EEDF in space and time due to the heating and cooling of electrons induced by the PSR is expected to strongly affect the radical densities and process rates in reactive gases. Furthermore, the enhanced control of the electron heating asymmetry might allow for an optimization of axial plasma density profiles, resulting in improvements of the total process rates and the lateral surface processing uniformity.

\ack

We thank James Franek (West Virginia University) and Uwe Czarnetzki (Ruhr-University Bochum) for helpful discussions. Funding by the Hungarian Scientific Research Fund through the grant OTKA K-105476 is gratefully acknowledged.

\end{document}